\documentclass[12pt,peerreview,draftcls]{IEEEtran}
\usepackage{mathrsfs}
\usepackage{algorithm}
\usepackage{algorithmic}
\usepackage{graphicx}
\usepackage{float}
\usepackage{hyperref}
\usepackage{cite}
\usepackage{color}
\usepackage{psfrag}
\usepackage{subfigure}
\usepackage{amssymb}
\usepackage{epsfig}
\usepackage{pifont}
\usepackage{amsmath}
\usepackage{array}
\usepackage{multicol}
\usepackage{multirow}
\usepackage{pifont}
\usepackage{indentfirst}
\usepackage{verbatim}
\usepackage{amsfonts}
\usepackage{amscd}
\usepackage{bm}
\usepackage{fancyhdr}
\usepackage{enumerate}

\newtheorem{proposition}{Proposition}

\IEEEoverridecommandlockouts

\newcommand{\apha}{{\alpha}}

\newcommand{\h}{{\|{\bf{h}}\|^2}}

\newcommand{\rou}{\delta_j}
\newcommand{\Pw}{P_{\max}}

\newcommand{\mapha}{\overline{\widehat\alpha_j}}
\newcommand{\dd}{\mathrm{d}}
\newcommand{\gp}{g_j^+}
\newcommand{\gm}{g_j^-}

\begin{document}
	
	\title{Exploiting Residual Resources to Support High Throughput with Resource Allocation }
	
	\author{		
		\IEEEauthorblockN{\large{Jia Guo, Chuting Yao, Chenyang Yang and Zixiang Xiong}} 
	}
	\maketitle
	\begin{abstract}
Residual radio resources are abundant in wireless networks due to dynamic traffic load, which can be exploited to support high throughput for serving non-real-time (NRT) traffic. In this paper, we investigate how to achieve this by resource allocation with predicted time-average rate, which can be obtained from predicted average residual bandwidth after serving real-time traffic and predicted average channel gains of NRT mobile users. We show the connection between the statistics of their prediction errors. We formulate an optimization problem to make a resource allocation plan within a prediction window for NRT users that randomly initiate requests, which aims to fully use residual resources with ensured quality of service (QoS). To show the benefit of knowing the contents to be requested and the request arrival time
in advance, we consider two types of NRT services, video on demand and video on reservation. The optimal solution is obtained, and an online policy is developed that can transmit according to the plan after instantaneous channel gains are available. Simulation and numerical results validate our analysis and show a dramatic gain of the proposed method in supporting high arrival rate of NRT requests with given tolerance on QoS.
	\end{abstract}
	
	\begin{IEEEkeywords}
		Predictive resource allocation, residual resource, high throughput, quality of service
	\end{IEEEkeywords}
	\vspace{-1mm}
	\section{Introduction}\vspace{-1mm}
To support the explosively growing traffic demands, various new techniques are under investigation for the fifth generation cellular networks and beyond \cite{UDNs2014Mag}. One of the main trends is continuing to provide higher spectral efficiency (SE), say by densifying the networks with more base stations (BSs) or more antennas. While further improving network SE is always beneficial, it has long been observed that the network resources are highly under-utilized \cite{EARTH-D4.1}. It has been recently observed from prevalent networks that in
average less than 15\% resource blocks are truly used in practice. One reason behind such a
dilemma is the temporal-spatial variation of traffic load, i.e., only some
BSs are busy during peak time of each day.


The dynamic nature of wireless traffic comes from user behavior, hence the traffic variation can be explored to boost network throughput by predicting the behavior.
While indeed random, human behavior exhibits strong regularity due to routine activity, as reported by big data analysis in a variety of disciplines \cite{science2010,froehlich2008route,Mardani2014Estimating,shi2014collaborative,bui2016anticipatory}. This implies the predictability of behavior-related information, either collectively or individually.
For example, the traffic volume and user trajectory are
predictable \cite{Mardani2014Estimating,WCNC17,nadembega2015destination}, from which future average resource usage status of a network and average channel gains of a user (with the help of a radio map \cite{KC16,RmapICC17}) can be derived \cite{Yao2017Data,abou2014toward}, and user preference can be predicted by machine learning such as  collaborative filtering \cite{shi2014collaborative}, from which the probability of a user requesting a content can be obtained.
As a consequence, predictive resource allocation is becoming one possible way to exploit residual resources \cite{Zheng2013Optimizing,Abou2013Predictive,abou2014toward,YTCOM2016}, which is applicable for both real-time (RT) and non-real-time (NRT) services \cite{bui2016anticipatory}.

\vspace{-4mm}\subsection{Related Works}
For RT traffic such as phone calls, predictive wireless access has been extensively investigated to improve the admission-level  quality of service (QoS), say reducing the call dropping rates during handover among adjacent cells \cite{Soh2006Predictive}. Considering that the information bits are generated randomly by each user and the RT service is with high priority, the major mechanism is to reserve resources for the RT traffic. Mobility prediction has long been used for mobility management to assist handover and for other location-based services, where the prediction granularity is in cell level or even more coarse (say, the next location) \cite{choi2002adaptive,bui2016anticipatory}. With the predicted next-cell connection and hand-off time, dynamical resource reservation and call admission control can be used to improve the QoS \cite{choi2002adaptive,nadembega2015mobility}.

For NRT traffic such as video on
demand (VoD) or file downloading, not only the admission-level and packet-level QoS of each user but also the performance
of a network can be improved by exploring future information. This is because the videos or files to be
transmitted is cacheable meanwhile the delay requirement of NRT traffic is not so stringent. As a result, the videos can be pre-buffered at a mobile station (MS)  when the MS is with good channel condition \cite{Abou2013Predictive} and/or is located in a cell with light traffic load \cite{YTCOM2016,Yao2017Data} (i.e., can be served with higher data rate \cite{Zheng2013Optimizing}). In contrast to non-predictive resource allocation that allocates radio resources at each time slot when instantaneous channel gain is available, predictive resource allocation makes a plan for assigning future resources in a prediction window at the start of the window when predicted information is available. The plan determines which BSs along the trajectory of a MS will serve the MS in which time slots with how much resources (say bandwidth).

Assuming that future instantaneous data rate in the prediction window is known, a resource allocation plan was optimized in \cite{Abou2013Predictive} to maximize the sum rate over the window, and a plan was made in  \cite{abou2014toward} to minimize the power consumption at BSs without causing stalling for VoD users. Because the instantaneous rate
is hard to predict, a more realistic
assumption is knowing the rate statistics in the future, say average data rate \cite{Zheng2013Optimizing} or data rate distribution
\cite{Nicola2014Model}.
Noticing that the rate prediction is
inevitably inaccurate even in average, a robust predictive resource allocation was proposed in \cite{Ramy2016JASC}, where the prediction errors on future rates are modelled as Gaussian noise.

\vspace{-4mm}\subsection{Motivation and Contributions}
All existing works implicitly assume that multiple NRT users initiate their requests simultaneously at the start of a prediction window. This assumption implies that the content to be requested and the exact request arrival time are known in advance, because the request arrivals are random and highly asynchronous in practice. However, only the probability of a content to be requested is predictable  \cite{shi2014collaborative} and the exact request arrival time is hard to predict if not impossible. As a consequence, it is unreasonable to assume knowing all future NRT request arrivals, unless the NRT users make reservations before truly requesting the videos or files as in video on reservation (VoR) \cite{veeravalli2006network}.

Besides, most priori research efforts assume that the future data rate is perfectly available or known with some statistics of prediction errors, but rarely address how the rate is predicted or how the error statistics are connected with the errors of predictable information.

Moreover, the time-varying rate is assumed only coming from large scale channel variation due to user mobility. This assumption implies that all radio resources can be used for NRT users. However, both RT and NRT requests may arrive in a cell, where the requests of RT users need to be served with higher priority and the requests of NRT users can be served with the residual resources after serving RT traffic. Therefore, the average rate of a NRT user depends not only on the trajectory but also on the variation of traffic load. This fact is largely overlooked in the literature of predictive resource allocation.


In this paper, we strive to demonstrate the performance gain of predictive resource allocation in supporting high throughput. To show the gain in real world networks, the request arrivals of NRT users are no longer assumed as synchronous. To show the benefit from knowing the contents to be conveyed and the request arrival time in advance, we consider two types of NRT services, VoD or VoR. We assume that average channel gains and average residual bandwidth are predictable from the traffic load and user trajectory prediction, by using the methods in \cite{Yao2017Data,abou2014toward}, with which the average rate prediction can be derived.  Since predicting user behavior is not an easy task, we show how the prediction errors of average rate are translated from those of predicted average channel gains and average residual bandwidth, and when it can be modelled as Gaussian as assumed in \cite{Ramy2016JASC}. Such analysis can help understand the gain from predicting different kinds of information and the required prediction accuracy to achieve the gain, which provides guidance for behavior prediction and facilitates robust optimization for predictive resource allocation.

The major contributions of this work are summarized as follows:
\begin{itemize}

\item{We  show the connection of the statistics of errors between the predicted average rate and the predicted average residual bandwidth and average channel gain, by resorting to the principle of maximum entropy.
We find that the prediction error of average rate mainly depends on the prediction error of average residual bandwidth, which implies that the user trajectory are unnecessary to be predicted accurately.}

\item{We formulate a problem to optimize resource allocation plan for randomly arrived NRT users that can exploit network residual resources in a prediction window. To maximize the request arrival rate of the
NRT users that the network can support and accommodate the uncertainty of requested content and request arrival time within the window, we minimize a weighted total transmission time with ensured maximal waiting time of the NRT users. We demonstrate the gain of the obtained optimal solution over priori solutions for predictive resource allocation by simulations.}


\end{itemize}

\emph{Notations:} $\|\cdot\|$ denotes Euclidean norm, and $|\cdot|$ denotes magnitude, ${\mathbb E}\{\cdot\}$ and ${\mathbb D}\{\cdot\}$ denote expectation and variance, $\mathbb N(\cdot)$ and $\mathbb U(\cdot)$ denote Gaussian and uniform  distributions, respectively.

The rest of the paper is organized as follows. In section II, we introduce channel and transmission models as well as a general traffic model with randomly arrived NRT requests. In section III, we analyze the prediction error statistics of average rate, formulate the resource allocation planning optimization problem, and find the optimal solution. In section IV, a transmission policy according to the plan is provided. Simulation and numerical results are shown  in section V, and the paper is concluded in section VI.

	\section{System Model}\vspace{-1mm}
	Consider a $N_b$ cell network, where each BS is equipped with $N_{\rm t}$ antennas, and serves two kinds of traffic with bandwidth $W_{\max}$ and transmit power $P_{\max}$. The first kind is RT traffic, and the other is NRT traffic. Because RT traffic has higher priority, the NRT traffic can be served by the residual resources of the network after the QoS of RT traffic is guaranteed. Given  dynamic traffic load of RT service, the residual resources available for NRT service is time-varying. For the MSs that request NRT traffic, we call them NRT users or simply MSs in the sequel.

Assume that there is a central processor (CP) in the network,
which makes the resource allocation plan for serving the NRT users within a prediction window.

	\subsection{Traffic and Channel Models}
		
The requests of NRT users arrive at the network randomly and asynchronously.
Each MS requests
 a video, either on-demand (i.e., VoD) or on reservation (i.e., VoR). For a MS demanding VoD service (called VoD MS), the CP can make the plan for resource allocation at the moment of the MS initiating its request. For the MS demanding VoR service (called VoR MS), the CP makes the plan at the moment of the MS making the reservation, which is earlier than the time instant that the MS starts to play the video. A video file is divided into multiple segments and then coded. Each segment is a stand-alone unit.  Once a segment is completely received by a MS, it can be decoded and played out. To avoid playback interruption due to empty playout buffers, a segment should be conveyed to the MS before the end of playing previous segment.

	Time is discretized into frames each with duration $\Delta$, and each frame includes $T_s$ time slots, each with duration of unit time (say 1 ms). The durations are defined according to the variation of large scale channel fading (including path-loss and shadowing) and small scale fading due to user mobility, respectively. Assume that the large scale channel gain (also called average channel gain) remains constant within each frame and may vary among frames, and the small scale channel gain (i.e., instantaneous channel gain, also called channel state information (CSI) in literature) remains constant within each time slot and varies among time slots with independent and identically distribution
(i.i.d.). For notational simplicity, we set the duration of the prediction window as $T_f$ frames and the playback duration for each segment as $T_{\rm seg}$ frames, and 	
we assume that each segment contains $B$ bits and each segment needs to play at the beginning of a frame.

\begin{figure}[!htb]
		\centering
		\vspace{-3mm}
        \subfigure[VoD traffic request model: MS$_K$ initiates a request at the start of a prediction window, i.e., the reference time.]{
		\includegraphics[width=0.92\textwidth]{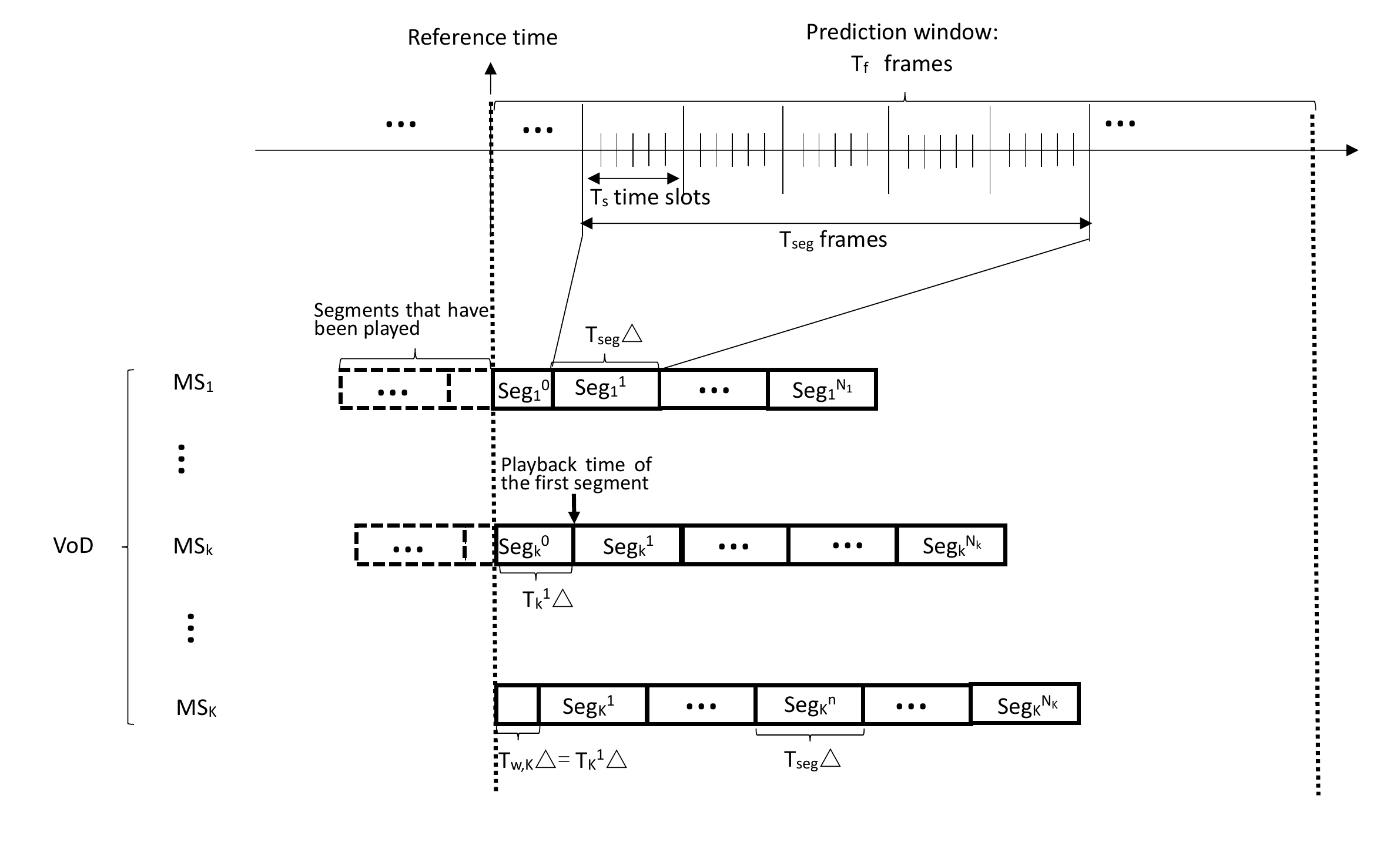}}\\
        \subfigure[VoR traffic request model: MS$_K$ makes a reservation at the start of a prediction window, and begins to play Seg$_K^1$ after a duration of $T_K^1 \Delta$.]{
		\includegraphics[width=0.92\textwidth]{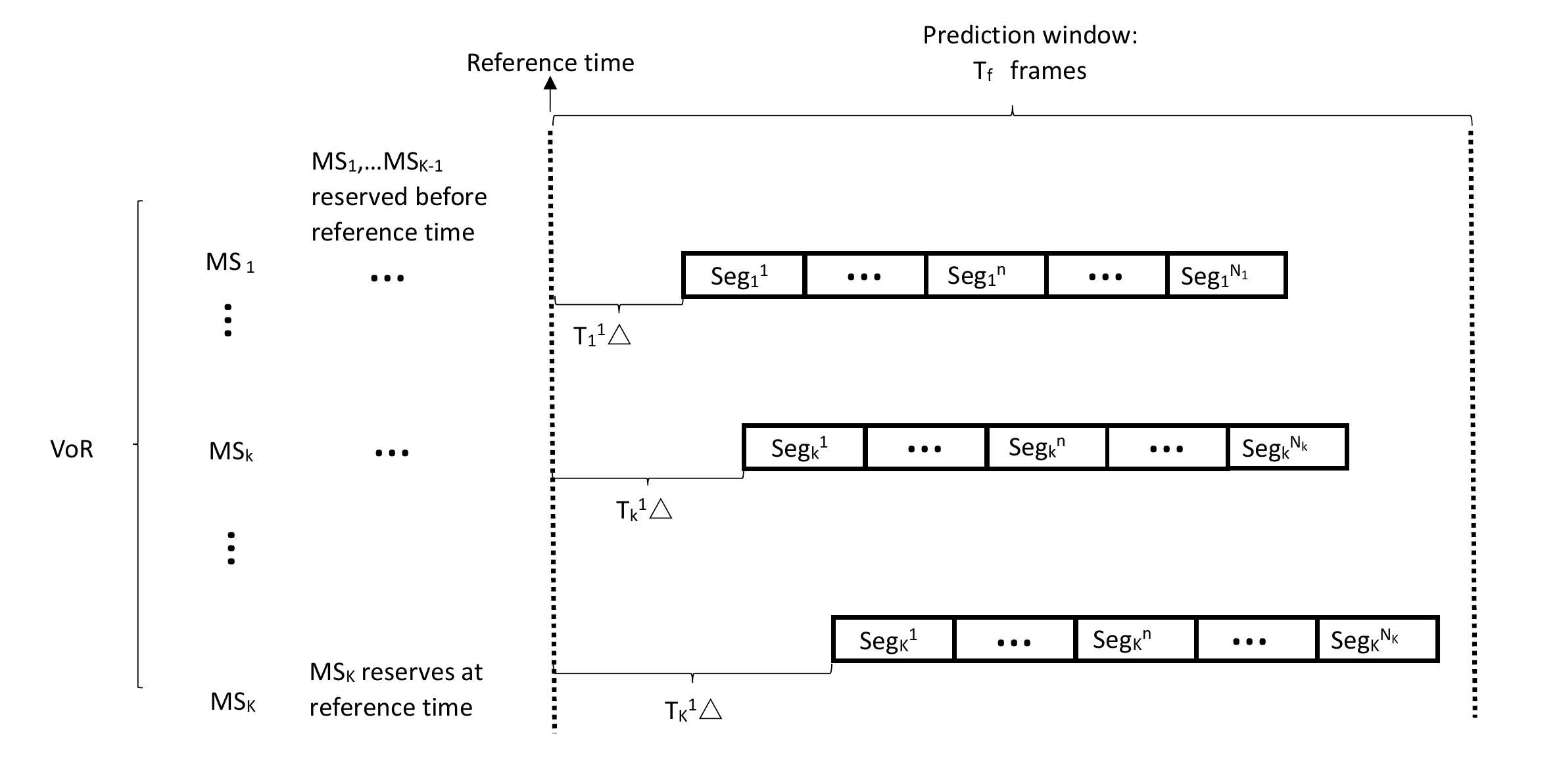}}\\
        \caption{Random request arrival model of NRT users. We set the request arrival time or reservation time of MS$_K$ as  the reference time. Before the reference time, MS$_1$ $\cdots$ MS$_{K-1}$ have sent requests or make reservations. After  the reference time, the requests or reservations of new MSs may arrive randomly in the window.}\label{reqmodel}
	\end{figure}

For the network only with VoD traffic in addition to RT traffic, we set the request arrival time of the $K$th MS (denoted as MS$_K$) as the start time of a prediction window, defined as the first time slot in the first frame (called \emph{reference time} for short). To reflect the random nature of the request arrivals, we consider the realistic scenario where $K-1$ VoD MSs are playing videos at the reference time, as shown in Fig. \ref{reqmodel}(a). This means that the prediction window is updated every time a new MS initiates a request. Within the window, new VoD MSs may initiate requests, whose arrival time is unknown at the reference time.

Denote the waiting time for MS$_k$ from the moment of sending a video request to the moment  of starting to play the video as $T_{{\rm w},k}$ frames, which reflects the initial delay. For VoD MSs, we can set $T_{{\rm w},k}\Delta$ as a constant duration that is long enough for downloading the first segment of a video (such as the advertisement time before the video being played). For the video requested by MS$_k$ who is playing a segment at the reference time (denoted as Seg$_k^0$), $N_k$ segments have not been played and wait to be downloaded within the window. Denote the  duration between the reference time and the moment of the first segment of MS$_k$ to be played in the window 
(denoted as Seg$_k^1$) as  $T_k^1 \Delta$.  For the $k$th ($k=1,\ldots,K-1$) VoD MSs, $T_k^1 \Delta$ is the residual playback duration of Seg$_k^0$, for the $K$th VoD MS, $T_k^1 \Delta$ is equal to its initial delay $T_{{\rm w},K}\Delta$. Denote the maximal waiting time a MS expected to watch the total video as  $T_{\rm mw}\Delta$, which is the sum of the initial delay and overall stalling time during playback \cite{seufert2015survey}. Then, $(T_{\rm mw}-T_{{\rm w},k})\Delta$ is the total stalling time allowed by MS$_k$, and hence $[{T_{\rm mw}-T_{{\rm w},k}+T_k^1+(n-1) T_{\rm seg}}]\Delta$ is the deadline for transmitting Seg$_k^n$, $n=1,\cdots, N_k$ without making MS$_k$ unsatisfied. Without loss of generality, assume that $T_{\rm seg}+T_{\rm mw} \leq T_f$.


For the network only with VoR traffic in addition to RT traffic, we set the time instant that MS$_K$ makes the reservation as the start time of the prediction window. At this moment, $K-1$ VoR MSs have already made the reservation, as shown in Fig. \ref{reqmodel}(b). For VoR MSs, the initial delay is $0$, i.e., $T_{{\rm w},k}=0$, $k=1,\ldots,K$.

	\subsection{Transmission Model}
To exploit residual resource, only the MS with highest average channel gain is associated with a BS, who serves the MS with all residual bandwidth and transmit power.
According to the resource allocation plan, there may exist multiple NRT users in each cell that should be served simultaneously. To avoid multi-user interference, various multiple access techniques can be
applied. For easy exposition, we consider time division multiple access, i.e., these MSs are served in different time slots. Then, maximal ratio transmission (MRT) is the optimal beamforming and hence the achievable rate of MS$_k$ in the $t$th time slot of the $j$th frame can be expressed as,
\begin{equation}\label{E:R_t}	
	R_{j,t}^k =  W_{j,t}\log_2\left(1+\frac{\alpha_j^k \|{\bf h}_{j,t}^k\|^2}{N_0 W_{j,t}} p_{j,t}\right),
\end{equation}
where $W_{j,t}$ and $p_{j,t}$ are respectively the residual bandwidth and transmit power in the $t$th time slot of the $j$th frame. In order to reflect the residual bandwidth after serving randomly arrived RT services with random service time, we model $W_{j,t}$ as i.i.d. random variables in all time slots of the $j$th frame \cite{Yao2017Data}. ${\bf{h}}^{k}_{j,t} \in \mathbb{C}^{N_t \times 1}$ is the small scale Rayleigh
fading channel vector with i.i.d. elements and ${\mathbb E}\{\|{\bf{h}}^{k}_{j,t}\|\}=N_t$, $\alpha_j^k$ is the large scale channel gain in the $j$th frame, and $N_0$ is the noise power spectrum density.
For easy analysis, assume that the residual transmit power is proportional to the residual bandwidth as in \cite{YTCOM2016}, i.e., $p_{j,t} = W_{j,t}P_{\max}/W_{\max}$.
Then, the time-average achievable rate in the $j$th frame (called \emph{average rate} for short) of MS$_k$ can be expressed as,
	\begin{equation}\label{R}	
	R_j^k = \frac{1}{T_s}\sum_{t=1}^{T_s} R_{j,t}^k = \frac{1}{T_s}\sum_{t=1}^{T_s} W_{j,t}\log_2(1+\frac{\alpha_j^k \|{\bf h}_{j,t}^k\|^2}{\sigma^2}P_{\max}),
	\end{equation}
where $\frac{\alpha_j^k \|{\bf h}_{j,t}^k\|^2}{\sigma^2}P_{\max}$ is instantaneous signal-to-noise ratio (SNR), and ${\sigma^2}=N_0 W_{\max}$.

	

	\section{Resource Allocation Planning with Predicted Information}\vspace{-1mm}
	In this section, we first show the connection between the statistics of prediction errors of the average rate and those of the average channel gain and residual bandwidth. Then, we formulate a resource allocation planning problem to use the residual resources for serving randomly arrived VoD MSs, and obtain the optimal solution. Finally, we extend the results to the network serving VoR MSs.
	
	\subsection{Statistics of Prediction Errors of Average Rate}
The small scale channel gain ${\bf h}_{j,t}^k$ is hard to predict beyond the channel coherence time, and the instantaneous residual bandwidth in each time slot $W_{j,t}$ is neither. As a result, the instantaneous data rate $R_{j,t}^k$ is hard to predict if not impossible. Fortunately, the trajectory of every NRT user and the traffic load of RT service at every BS are predictable within the prediction window  \cite{Mardani2014Estimating,WCNC17,nadembega2015destination}. Then, the CP can predict the average channel gains in each frame for each MS with the help of a radio map \cite{Abou2013Predictive}, as well as the average residual bandwidth in each frame at each BS with the predicted traffic load \cite{Yao2017Data}. In practice, the prediction is never perfect. Denote the predicted residual bandwidth in the $j$th frame as $\widehat{W_j}$, which is with mean value of $\overline{\widehat{W_j}}$ and variance $\sigma_{\widehat{W_j}}^2$.
Denote the predicted large scale channel gain for MS$_k$ in the $j$th frame as $\widehat {\alpha_j^k}$, which is with mean value of $\overline{\widehat{\alpha_j^k}}$ and bounded uncertainty of $\delta_j^k/2$ (i.e., $\overline{\widehat{\alpha_j^k}}-\delta_j^k/2 \leq \widehat {\alpha_j^k}\leq \overline{\widehat{\alpha_j^k}}+\delta_j^k/2$).


By using the predicted residual bandwidth in each frame as the residual bandwidth in each time slot and using the predicted average channel gain, and considering that ${\bf h}_{j,t}^k$ is i.i.d., if $T_s\to \infty$, then from \eqref{E:R_t} and \eqref{R} we can express the predicted time-average rate as,
	\begin{eqnarray}\label{aveR}
	\widehat{R_j^k}=\frac{1}{T_s}\sum_{t=1}^{T_s}\widehat{W_j}\log_2\Big(1+\frac{\widehat{\alpha_j^k}\|{\bf{h}}^k_{j,t}\|^2}{\sigma^2}P_{\max}\Big)\!=\!\widehat{W_j}\mathbb E\Big\{\log_2\Big(1+\frac{\widehat{\alpha_j^k}\|{\bf{h}}^k_{j,t}\|^2}{\sigma^2}P_{\max}\Big)\Big\},
	\end{eqnarray}
where the average is taken over small scale channel.
	
	For a random variable $X$, the expectation of its function $\varphi( X)$ can be approximated as
	\cite{approx}
	\begin{equation}
	\label{equ.1_approx}
	\mathbb{E}\{\varphi( X)\} = \mathbb{E}\{\varphi(\mu_x +  X - \mu_x )\}\approx \mathbb{E}\{\varphi(\mu_x) + \varphi'(\mu_x)( X-\mu_x)\}= \varphi(\mu_x),
	\end{equation}
	where $\mu_x = \mathbb{E}\{ X\}$, and the approximation is accurate when the variance of $X$ is small. With this approximation and $\mathbb E\{\|{\bf h}_{j,t}^k\|^2\}=N_{t}$, \eqref{aveR} can be approximated as,
	\begin{eqnarray}\label{meanRj1}
	\widehat{R_j^k}
\approx\widehat{W_j}\log_2\Big(1+\frac{\widehat{\alpha_j^k}N_t }{\sigma^2}P_{\max}\Big),
	\end{eqnarray}
	which is accurate when $N_t$ is large.

The prediction errors of $\widehat{W_j}$ and $\widehat {\alpha_j^k}$ depend on the prediction algorithms of traffic load and user trajectory as well as the interpolation algorithms to derive the fine-grained average residual bandwidth and average channel gain from a coarse-grained prediction and radio map construction. There is no model available for the distribution of $\widehat{W_j}$ and $\widehat {\alpha_j^k}$ in the literature that are validated by viable algorithms on real data trace. To gain some useful insight, we model the predictions according to the \emph{principle of maximum entropy} \cite{Jaynes1957Information}. With given mean value and variance, Gaussian distribution is with maximum entropy, and with given upper and lower bounds,  uniform distribution is with maximum entropy \cite{Park2009Maximum}. Since the mean value and variance of the prediction of traffic load (and hence residual bandwidth) could be obtained \cite{WCNC17}, the predicted average residual bandwidth can be modelled as Gaussian distribution. Since user trajectory in a short horizon is bounded by road topology \cite{nadembega2015destination} and shadowing can be approximated as bounded, we model the predicted average channel gain as uniform distribution.
Then, the following proposition shows how the statistics of the prediction errors of average residual bandwidth and average channel gain translate to
the statistics of the prediction errors of average data rate. Such a relation can provide a design guidance for the required accuracy on predicting average residual bandwidth and average channel gain.

\begin{proposition}\label{Pro:2}
	If (i) $T_s\to\infty$, (ii) $\widehat{W_j}\thicksim{\mathbb N}(\overline{\widehat{W_j}},\sigma_{\widehat{W_j}}^2)$, (iii) $\widehat{\alpha_j^k}\thicksim{\mathbb U}(\overline{\widehat{\alpha_j^k}}-\delta_j^k/2,\overline{\widehat{\alpha_j^k}}+\delta_j^k/2)$,
(iv) the predicted average and instantaneous  SNRs are large and $\delta_j^k\ll\overline{\widehat{\alpha_j^k}}$,
then the average rate prediction error $\widetilde{R_j^k} =\widehat{R_j^k} - R_j^k$ follows Gaussian distribution, which has mean value
	\begin{align}\label{E:average_R}
	\overline{\widetilde{R_j^k}} \approx \overline{\widehat{W_j}}\widehat{\mu^k_{j}}-\overline{W_j}\Bigg(\log_2\Big(\frac{{\alpha_j^k}}{\sigma^2}P_{\max}\Big)+\frac{\psi(N_t)}{\ln 2}\Bigg),
	\end{align}
	and variance
	\begin{align}\label{E:Var_R}
	\sigma_{\widetilde{R_j}}^2 \approx (\sigma_{\widehat{W_j}}^2+\overline{\widehat{W_j}}^2)(\widehat{\sigma_j^k}^2+\widehat{\mu_j^k}^2) - \overline{\widehat{W_j}}^2\widehat{\mu^k_{j}}^2,
	\end{align}\
	where
	\begin{subequations}\label{E:xi_and_lambda}
		\begin{align}
		&	\widehat{\sigma_{j}^k}^2 \approx  \frac{1}{{\delta_j^k}^2\ln^22}\Bigg(\Big(\frac{{\delta_j^k}^2}{4}-{\overline{\widehat{\alpha_j^k}}}^2\Big)\ln^2\frac{{\overline{\widehat{\alpha_j^k}}}+{\delta_j^k}/2}{{\overline{\widehat{\alpha_j^k}}}-{\delta_j^k}/2}+{\delta_j^k}^2\Bigg),
		\label{xia}\\
		&	\widehat{\mu_j^k}\approx \frac{1}{{\delta_j^k}\ln2}\Big({\overline{\widehat{\alpha_j^k}}}\ln\Big(\frac{{\overline{\widehat{\alpha_j^k}}}+{\delta_j^k}/2}{{\overline{\widehat{\alpha_j^k}}}-{\delta_j^k}/2}\Big)+\frac{{\delta_j^k}}{2}\ln\frac{({\overline{\widehat{\alpha_j^k}}}^2-{\delta_j^k}^2/4)P_{\max}^2}{\sigma^4}
		+{\delta_j^k}(\psi(N_{\rm t})-1)\Big) \label{xib},
		\end{align}
	\end{subequations}
 $\overline{W_j}={\mathbb E}\{W_{j,t}\}$, $\psi(\cdot)$ is the Euler's digamma function, $\psi'(\cdot)$ is the derivative of $\psi(\cdot)$. When $\widehat{W_j}$ or $\widehat{\alpha_j^k}$ is biased, the impact of the prediction bias of large scale channel gain is much smaller than that of the residual bandwidth on the prediction bias of average rate.
\end{proposition}
\begin{IEEEproof}
See Appendix A.
\end{IEEEproof}
\vspace{1mm}	

Later simulations show that the results in Proposition 1 still hold when $\widehat {\alpha_j^k}$ is Gaussian, $T_s$ is not so large, the values of $\delta_j^k$ and $\overline{\widehat{\alpha_j^k}}$ are comparable, and the SNRs are not high.

\vspace{-2mm}	
\subsection{Optimizing the Resource Allocation Plan for the VoD MSs}
	At the beginning  of the prediction window, the CP can make a resource allocation plan for serving the NRT users with the predicted time-average rates. To achieve the goal of fully using the residual resources for supporting high throughput of NRT users, we optimize the plan (i.e., the time resources allocated to the VoD MSs) denoted as $[{\bf s}^1,\ldots,{\bf s}^K]$, where ${\bf s}^k=[s_1^k,\ldots,s_{T_f}^k]^T$, and $s^k_j\in[0,1]$ is the percentage of the time slots assigned to MS$_k$ in the $j$th frame.
	
	Denote the objective function as $f({\bf s}^1,\ldots,{\bf s}^K)$.
To maximize the arrival rate (i.e., throughput) of the NRT users that the network can support, one way is to directly maximize the amount of data transmitted during the prediction window (equivalent to maximize the sum rate over the window \cite{Abou2013Predictive}) or to indirectly minimize the total transmission time, each with ensured QoS  \cite{abou2014toward}. Yet such objectives cannot exploit residual resources in the network with randomly arrived VoD requests.
To help understand how to find a proper objective function to achieve our goal, we first analyze the behavior of the policies optimized toward these two objectives in a special case: there is only one VoD MS in the network, who requests only one segment (i.e., $N_1=1$) at the reference time. Then, the playback duration is $T_{\rm seg}$ frames, and the QoS is to complete the transmission for the $B$ bits before the playback of the segment.

In this special case, the problem that maximizes the overall amount of data transmitted over the prediction window meanwhile ensures no stalling for the VoD MS can be simplified as,
	\begin{subequations}\label{P:Ratemax1MS}
		\begin{align}
		\max_{{\bf s}^1} & \sum_{j=1}^{T_f} s_{j}^1 \widehat{R_j^1}  
		\\ s.t. & \sum_{j=1}^{T_{\rm seg}} s_{j}^1 \widehat{R_j^1} \Delta = B, \label{P:Ratemax1MS-C1}
		\\ &0 \leq  s_{j}^1 \leq 1, j=1,\ldots,T_{f}, \label{P:Ratemax1MS-C2}
		\end{align}
	\end{subequations}
where $f({\bf s}^1)=\sum_{j=1}^{T_f} s_{j}^1 \widehat{R_j^1} $, and $\Delta$ is a constant and hence is removed from the objective function. It is easy to find that if the BS can transmit $B$ bits to the MS during $T_{\rm seg}$ frames, the optimal solution is any vector ${\bf s}^1$ satisfying $\sum_{j=1}^{T_{\rm seg}} s_{j}^1 \widehat{R_j^1} \Delta= B$ and \eqref{P:Ratemax1MS-C2}, which is not unique.\footnote{When $N_1>1$, the optimal solution of this problem (i.e., the allocated resources to transmit all the $N_1$ segments) is still not unique. This is because the QoS constraint becomes $\sum_{j=1}^{T_{\rm seg}} s_{j}^1 \widehat{R_j^1} \Delta > B$ for the first $N_1-1$ segments, while the constraint in \eqref{P:Ratemax1MS-C1} should be satisfied for the last segment of the video. } In this case where the residual resource in the BS is sufficient to convey the $B$ bits, the objective function is no use at all, because there are only $B$ bits required to transmit in the window. Otherwise, if the constraint in \eqref{P:Ratemax1MS-C1} cannot be satisfied, the problem is infeasible. In this case where the residual resource is insufficient for ensuring the QoS of the VoD MS, a simple technique is to use all residual resources in $T_{\rm seg}$ frames for transmission. This suggests
that such a formulation is not appropriate to optimize predictive resource allocation for the network with residual resources.
	
	In the special case, another problem that minimizes the total transmission time in the window meanwhile ensures
no stalling can be simplified as
	\begin{subequations}\label{P:context1MS}
		\begin{align}
		\min_{{\bf s}^1} & \sum_{j=1}^{T_f} s^1_j  \label{P:context1MS-O}
		\\ s.t. & \sum_{j=1}^{T_{\rm seg}} s_{j}^1 \widehat{R_j^1} \Delta = B, \label{P:context1MS-C1}
		\\ &0 \leq  s_{j}^1 \leq 1, j=1,\ldots,T_{f}, \label{P:context1MS-C2}
		\end{align}
	\end{subequations}
where $f({\bf s}^1)=\sum_{j=1}^{T_f} s^1_j $, and again $\Delta$ is removed from the objective function.

We can see that if both problems \eqref{P:Ratemax1MS} and \eqref{P:context1MS} are feasible, then the optimal solution of problem \eqref{P:context1MS} is one  of the solution of problem \eqref{P:Ratemax1MS} that minimizes the total time for transmission.
Problem \eqref{P:context1MS} is a linear programming, which can be solved by the simplex problem. If the problem is feasible, the solution can be expressed as,
	\begin{eqnarray}\label{S:P:context1MS}
	s_{j_{i}}^{1*}=\left\{
	\begin{aligned}
	& \max\Big(\min\Big(\frac{B-\sum_{m=0}^{i-1} \widehat{R_{{j_m}}^1}s_{j_m}^{1*}\Delta}{\widehat{R_{j_{i}}^1}\Delta},1\Big),0\Big), & 1\leq j_i \leq T_{\rm seg}, ~i \geq1 \\
	& 0, & j_i > T_{\rm seg}
	\end{aligned} \right.
	\end{eqnarray}
	where $\widehat{R_{j_1}^1},\cdots,\widehat{R_{j_{T_{\rm seg}}}^1}$ are the descending ordered $\widehat{R_{1}^1},\cdots,\widehat{R_{T_{\rm seg}}^1}$.
It can be seen that the CP always sequentially selects the frames in the window with the largest achievable rates for transmission.

Now, we come back to the general problem with multiple MSs each with multiple segments.
  In practice, a new request for VoD traffic may arrive in the prediction window, but the arrival time is hard to know at the reference time. With the solution of problem \eqref{P:context1MS}, when the new VoD MS arrives, some VoD MSs whose requests already arrive at the reference time (e.g., one or more MSs among MS$_1 \cdots$ MS$_{K-1}$ in Fig. \ref{reqmodel}) may not have received any bits due to still not experiencing the best channels. Then, the VoD MSs may compete for the remaining time resources in the window, and the resources before the new MS arrives is wasted.

  Inspired by the observation from the analysis on the special case, we introduce an alternative objective function.
  To fully use the residual resources under the uncertainty on future arrived requests, the data of the arrived VoD MSs should be transmitted in the earlier frames that are closer to the reference time. A natural way to employ more time slots in the early frames is to define the objective function for multiple MSs as
	$f({\bf s}^1,\ldots,{\bf s}^K)= \sum_{j=1}^{T_f} \sum_{k=1}^K \omega(j) s^k_j $,
where the weighting function $\omega(j)$ should increase with $j$. To balance the usage of the early frames close to the reference time and those with higher rate, we can simply set $\omega(j)=j$ as an illustration. We can also select other weighting functions, which do not change the optimization problem and achieve similar performance.

To control the QoS of the VoD MSs, we impose constraint on the maximal waiting time for each MS to watch the total video, $T_{\rm mw}\Delta$, which is the sum of the initial delay and overall time of stalling during playback. Then, the expected deadline of  MS$_k$ for transmitting all required $\sum_{i=1}^{n} B_k^i$ bits to play Seg$_k^n$ is $[{T_{\rm mw}-T_{{\rm w},k}+T_k^1+(n-1) T_{\rm seg}}]\Delta$, $n=1,\cdots, N_k$.

	For MS$_k$, there are $N_k$ segments to be played, and the playback duration of each segment is $T_{\rm seg}\Delta$.
To exploit the resources in the network and guarantee the QoS of the $K$ MSs whose requests have arrived at the reference time, the resource planning problem is formulated as,
\begin{subequations}\label{P1}
\begin{align}
	{\bf P1}:\min_{T_{\rm mw}, {\bf s}^1,\ldots,{\bf s}^K} & \sum_{j=1}^{T_f} \sum_{k=1}^K j\cdot s^k_j \\
s.t. &~\sum_{j=1}^{T_{\rm mw}-T_{{\rm w},k}+T_k^1+(n-1) T_{\rm seg}}  s^k_{j} \widehat{R^k_j}\Delta\geq \sum_{i=1}^{n}B_k^i,n=1,\dots,N_k-1,\label{P1-b}\\
&\sum_{j=1}^{T_{\rm mw}-T_{{\rm w},k}+T_k^1+(N_k-1) T_{\rm seg}}  s^k_{j} \widehat{R^k_j}\Delta=\sum_{i=1}^{N_k}B_k^i,\label{P1-c}\\
&\sum_{k\in{\cal K}_{j,i}} s^k_j\leq 1,j=1,\ldots,T_f, i=1,\ldots,N_b,\label{P1-d}\\
&s^k_j\in[0,1],j=1,\ldots,T_f,~k=1,\ldots,K,\label{P1-e}
	\end{align}
\end{subequations}
where \eqref{P1-b} and \eqref{P1-c} are the QoS constraints, \eqref{P1-d} is the total resource constraint at the $i$th BS, and ${\cal K}_{j,i}$ is the set of MSs in the coverage of the $i$th BS in the $j$th frame.	

	Problem $\bf P1$ has two kinds of variables, the first is the maximal waiting time $T_{\rm mw}$, and the second is the resource planning vector ${\bf s}_j^k$. When the value of $T_{\rm mw}$ is fixed, the problem reduces to a linear programming \cite{schrijver1998theory} as follows,
	\begin{align}\nonumber
	{\bf P2}:\min_{{\bf s}^1,\ldots,{\bf s}^K} & f( {\bf s}^1,\ldots,{\bf s}^K)
	\\ s.t. &~\eqref{P1-b},\eqref{P1-c},\eqref{P1-d},\eqref{P1-e}, ~k=1,\ldots,K,
	\end{align}
	since \eqref{P1-d} and \eqref{P1-e} become linear constraints of variables $s^k_j$. Then, problem ${\bf P2}$ can be easily solved if the problem is feasible.
	
	When $T_{\rm mw}$ decreases, the feasible region of problem ${\bf P2}$ reduces. The minimal value of $T_{\rm mw}$ to make problem ${\bf P2}$ feasible can be found by bisection searching, which is denoted as $T_{\rm{mw}}^*$. Given this value of $T_{\rm{mw}}^*$, the optimal resources assigned to the $K$ MSs can be obtained as ${\bf s}^{k*}=[s^{k*}_1,\ldots,s^{k*}_{T_f}]^H$, which is the global optimal solution of problem $\bf P1$.

{\emph{Remark}}:	At the reference time when MS$_1, \cdots,$ MS$_{K-1}$ have sent their requests and MS$_K$ initiates its video request, the resource allocation plan is made for all the $K$ MSs by solving problem $\bf P1$. The CP needs to re-make a plan in the following scenarios: (i) when a new MS initiates a request. In this case, the CP re-makes the plan for all MSs (including the new MS) in the network; (ii) when a prediction window finishes before all segments requested by existing MSs are downloaded. In this case, CP re-makes a plan for transmitting the residual segments.

    \subsection{Optimizing the Resource Allocation Plan for VoR MSs}
Similar problem can be formulated for VoR MSs. When $K-1$ VoR MSs have already made their reservation before the reference time and a VoR MS makes its reservation at the reference time, the only difference between the VoR MSs and VoD MSs lies in that the initial delay is zero, i.e., $T_{{\rm w},k}=0$.
Then, a simplified problem from problem $\bf P1$ can be obtained. Again, a re-plan can be made similar to the system with VoD MSs.

	\section{Transmission Policy According to Resource Allocation Plan}
With the resource allocation plan ${\bf s}^{k*}=[s^{k*}_1,\ldots,s^{k*}_{T_f}]^H$, which MS should be served by (and hence associated with) which BS along the MS's trajectory can be determined. At the start of each time slot, small scale channel vector of each MS can be estimated at its associated BS. Since more than one MS may be associated with a BS, user scheduling is necessary at each time slot. To maximize the number of satisfied MSs (i.e., the NRT users whose video files are completely
	conveyed before their expected deadline), the BS schedules the MSs according to their \emph{transmission progress}, defined as
	\begin{align}
	\Lambda (k,J)= \sum_{j=1}^J s^{k*}_{j} \widehat{R^k_{j}}\Delta,
	\end{align}	
which is the amount of data ought to be accumulatively conveyed at the end of the $J$th frame ($J=1,\cdots,T_f$). It can be computed by the CP at the start of the prediction window after making the resource allocation plan.
%
%

	In the $t$th time slot of the $J$th frame, the set of MSs who are planned to be served by the $i$th BS but have not caught up the transmission progress can be expressed as
	\begin{align}
\tilde {\cal K}_{J,i}\triangleq \{k\in{\cal K}_{J,i}|\Lambda(k,J)-\frac{\Delta}{T_s}(\sum_{l=1}^{j-1} \sum_{\tau=1}^{T_s} R^{k}_{l,\tau}+ \sum_{\tau=1}^{t-1}R^{k}_{j,\tau})>0\}.
\end{align}
To exploit the residual resources, the $i$th BS selects the MS with  maximal instantaneous achievable rate from this MS set, i.e., according to the following rule \vspace{-1mm}
	\begin{equation}
	\textstyle k^* = \arg \max_{k} \{R^k_{j,t}|s^{k*}_{j}>0 \text {~and~} k\in \tilde {\cal K}_{J,i} \}. \vspace{-1mm}
	\end{equation}
	Then, the $i$th BS serves the $k^*$th MS with MRT using the instantaneous residual
	transmit power and residual bandwidth $W_{j,t}$ and $p_{j,t}$.
	
	Due to the prediction error on the time-average rate, it may happen that some MSs do not catch up the transmission progress at the end of a frame. In this case, the BS  transmits the remaining data to these MSs at the beginning of the next frame, no matter if other segments need to be transmitted in the frame. After the remaining data have been conveyed, the BSs start to transmit the segments according to the plan. Despite that such a strategy may cause a ``mismatch" between actual transmission progress and the planned progress, the mismatch can be controlled by the re-plan mechanism.

	\section{Simulation and Numerical Results}
	In this section, we validate previous analysis via numerical results and demonstrate the performance gain of predictive resource allocation by simulations.
	
	\subsection{Simulation Set-Up}
\begin{figure}[!htb]
		\centering
		\vspace{-3mm}
		\includegraphics[width=1\textwidth]{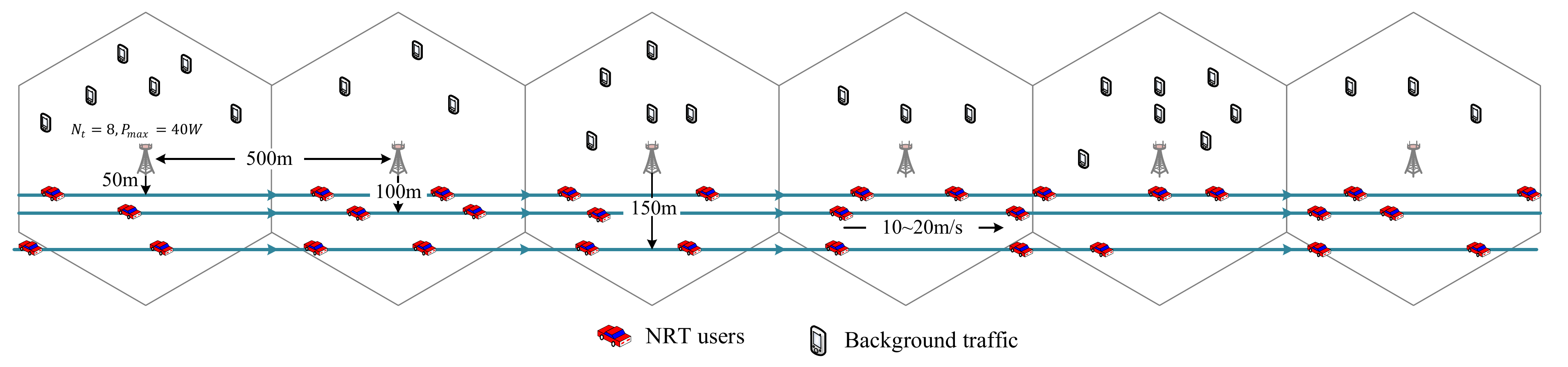}\\
		\caption{System setup in simulation.}\label{traffic}\vspace{-10mm}
	\end{figure}

	Consider a cellular network with six BSs, each equipped with $N_t=8$ antennas, which are located along a straight line. The cell radius is $D=250$ m. As shown in Fig. \ref{traffic}, the NRT users  move along three roads of straight lines with minimum distance from the BSs as $50$ m, $100$ m and $150$ m, respectively.  Each MS requests a video with size of $B=20$ Mbytes and playback duration of $100$ s. Each video consists of $N=10$ segments, i.e., each segment with size of $2$ Mbytes is played out for $T_{\rm seg}=10$ s.  The prediction window contains $T_f=300$ frames. Each frame is with duration of one second, and each time slot is with duration $\Delta=10$ ms, i.e., each frame contains $T_s=100$ time slots (which is far from infinity as we assumed in analysis).

	The video requests of the MSs randomly arrive only between the $1$st frame and $100$th frame in the prediction window (when they arrive uniformly within the 300 frames, the results are similar). To characterize the different resource usage status of the BSs in serving the RT traffic in an under-utilized network, we consider two types of BSs:  busy BS with average residual bandwidth in each frame (say the $j$th frame) as $\overline{W_j}=1$ MHz and idle BS with $\overline{W_j}=10$ MHz, which are alternately located along the line as idle, busy, idle, busy, idle, and busy BS. Considering that the prediction error of traffic load is within $20$\% as reported in \cite{Mardani2014Estimating}, the predicted average residual bandwidth changes among frames according to $\widehat{W_j}\thicksim\mathbb{N}(\overline{\widehat{W_j}},\sigma_{\widehat{W_{j}}}^2)$, where $\sigma_{\widehat{W_{j}}}/\overline{\widehat{W_j}}=0.2$. To reflect the prediction error of user trajectory, the predicted large scale fading gains vary  among frames according to $\widehat{\alpha_j^k}\thicksim \mathbb{U} (\overline{\widehat{\alpha_j^k}}-\delta_j^k/2,\overline{\widehat{\alpha_j^k}}+\delta_j^k/2)$, where $\delta_j^k/\overline{\widehat{\alpha_j^k}}=1$, which corresponds to the variation range of path loss between cell center and cell edge. We consider unbiased prediction for $\widehat{W_j}$ and $\widehat{\alpha_j^k}$, i.e., $\overline{\widehat{W_j}}=\overline{W_j}$ and $\overline{\widehat{\alpha_j^k}}=\alpha_j^k$.

The maximal transmit power of each BS is 40 W and cell-edge SNR is set as 5 dB, where the intercell interference is implicitly reflected. Since shadowing has little impact on the performance, we only consider path loss in average channel gain to reduce the time for simulation. The path loss model is $36.8 + 36.7\log_{10}(d)$, where $d$ is the distance between the BS and MS in meter. The results are obtained from 100 Monte Carlo trails. In each trail, the trajectory, request arrival and channel gain of each MS change randomly. In particular, for each MS, the moving speed is uniformly distributed in $(10,20)$ m/s, the moving direction is uniformly selected
as -180 or +180 degree,  and the location where the MS initiates a request is randomly selected from the three roads. The requests of the MSs arrive from the $1$st to the $100$th frame according to Poisson process with given average arrival rate. Besides, the small-scale channel in each time slot changes independently according to Rayleigh fading. This setup will be used in the following simulation, unless otherwise specified.
	
	\subsection{Resource Allocation Schemes for Comparison and Evaluation Metrics}
	We consider several resource allocation schemes for comparison, which can be divided into two categories of predictive and non-predictive schemes. With predictive schemes, the CP can make resource allocation plan with the predicted time-average rate in \eqref{meanRj1}, while with non-predictive schemes, the CP does not predict any information, as listed in the following.\\
\emph{\textbf{Predictive schemes:}}
	\begin{itemize}
	\item \textbf{Proposed}: The resource allocation plan is found from the solution of problem ${\bf P2}$, and the transmission policy in section IV is used.
	\item \textbf{Max-Throughput}: The resource allocation plan is made to maximize the time-average sum rate over the prediction window under the constraints in \eqref{P1-b}-\eqref{P1-e} (the optimization problem degenerates into problem \eqref{P:Ratemax1MS} when there is only one MS and the video is only with one segment), which has the same objective function as the method proposed in \cite{Abou2013Predictive}. Since the optimal solution is not unique, we can use any solution found from the constraints.
	\item \textbf{Min-Time}: This is the method proposed in \cite{abou2014toward}, where the resource allocation plan is made to minimize the total transmission time of all MSs in the prediction window (the optimization problem degenerates into problem \eqref{P:context1MS} when there is only one MS  and the video is only with one segment).
	\end{itemize}
\emph{\textbf{Non-predictive schemes:}}
	\begin{itemize}
	\item \textbf{Non-predictive w/o QoS}: Each BS serves all MSs with best effort. In each time slot, the BS only serves the MS with the highest instantaneous data rate.
	\item \textbf{Non-predictive w QoS}: This is the scheme proposed in \cite{su2015user}, where each BS serves the MS with the earliest deadline in each time slot. If several MSs have the same deadline, then the MS with most bits to transmit is served first.
	\end{itemize}

	We consider two performance metrics: the average stalling time of all MSs and the maximal request arrival rate of MSs when the maximal stalling time expected by 99.9\% of the MSs are satisfied. The first metric measures the QoS of the VoD MSs. The second metric measures the traffic carrying ability of the network for supporting the MSs with given tolerance on QoS. Other metrics such as stalling frequency are also used to evaluate the QoS in the sequel.
	
	\subsection{Simulation and Numerical Results}
	\subsubsection{Validating the analysis} We first validate the proposition.

We consider a typical scenario where MS$_k$ is served by a busy BS at the $j$th frame, i.e., $\overline{W_j}=1$ MHz, and $\sigma_{W_j}$ is set as $0.2$ MHz. To reflect the uncertainty of prediction, we set $\sigma_{\widehat{W_j}}$ as $0.2$ MHz and $\delta_j^k/\alpha_j^k=1$. The results for other settings are similar, and hence are not shown.

\begin{figure}[!htb]
	\begin{minipage}[t]{0.5\linewidth}	
	\centering
	\vspace{-3mm}
	\subfigure[Numerical and simulated PDF of average rate.]{
		\includegraphics[width=\textwidth]{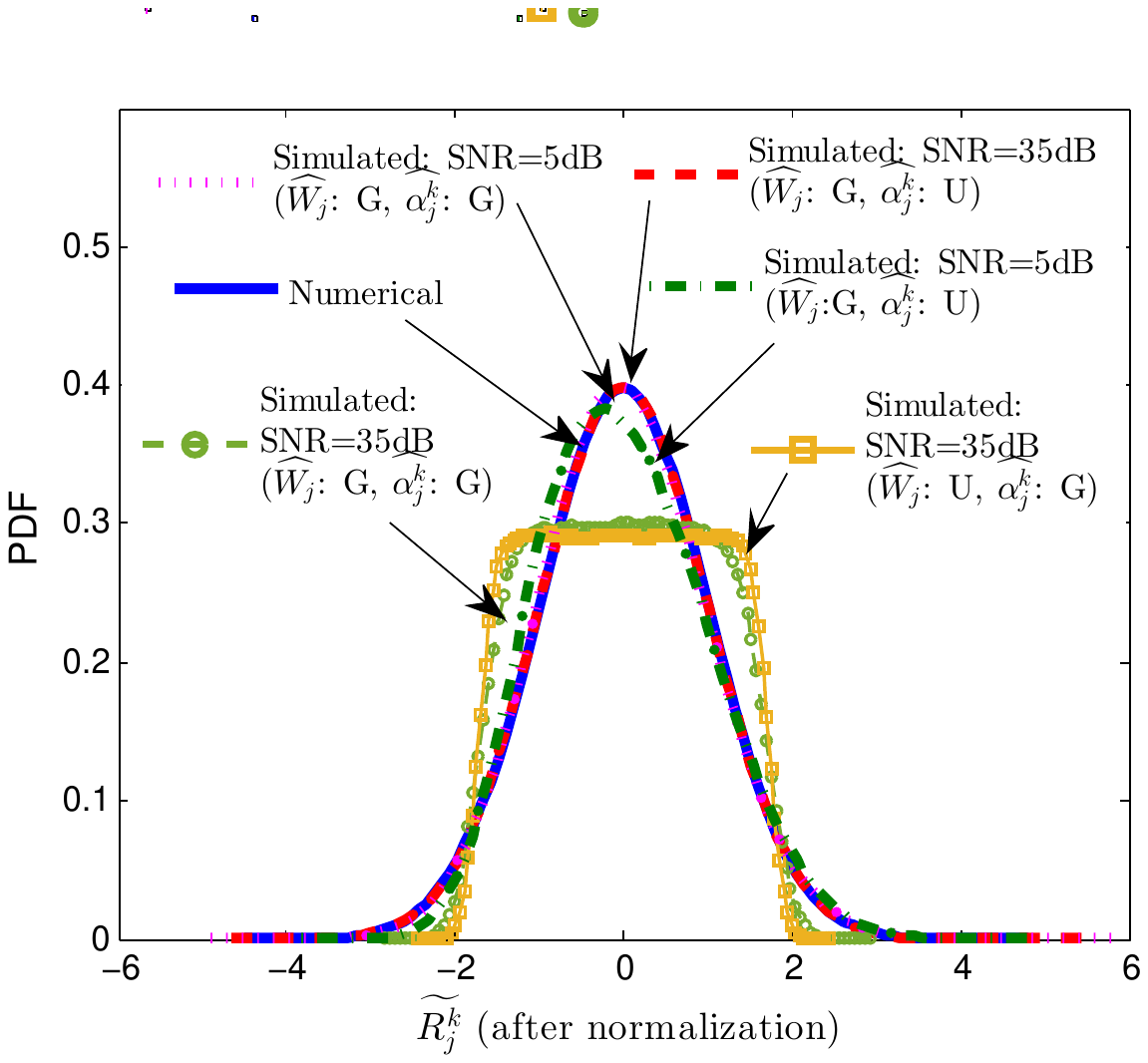}}
	\end{minipage}
	\begin{minipage}[t]{0.5\linewidth}	
	\subfigure[Normalized approximation errors of $\overline{\widetilde{R_j^k}}$ and $\sigma_{\widetilde{R_j^k}}$ versus SNR.]{
		\includegraphics[width=\textwidth]{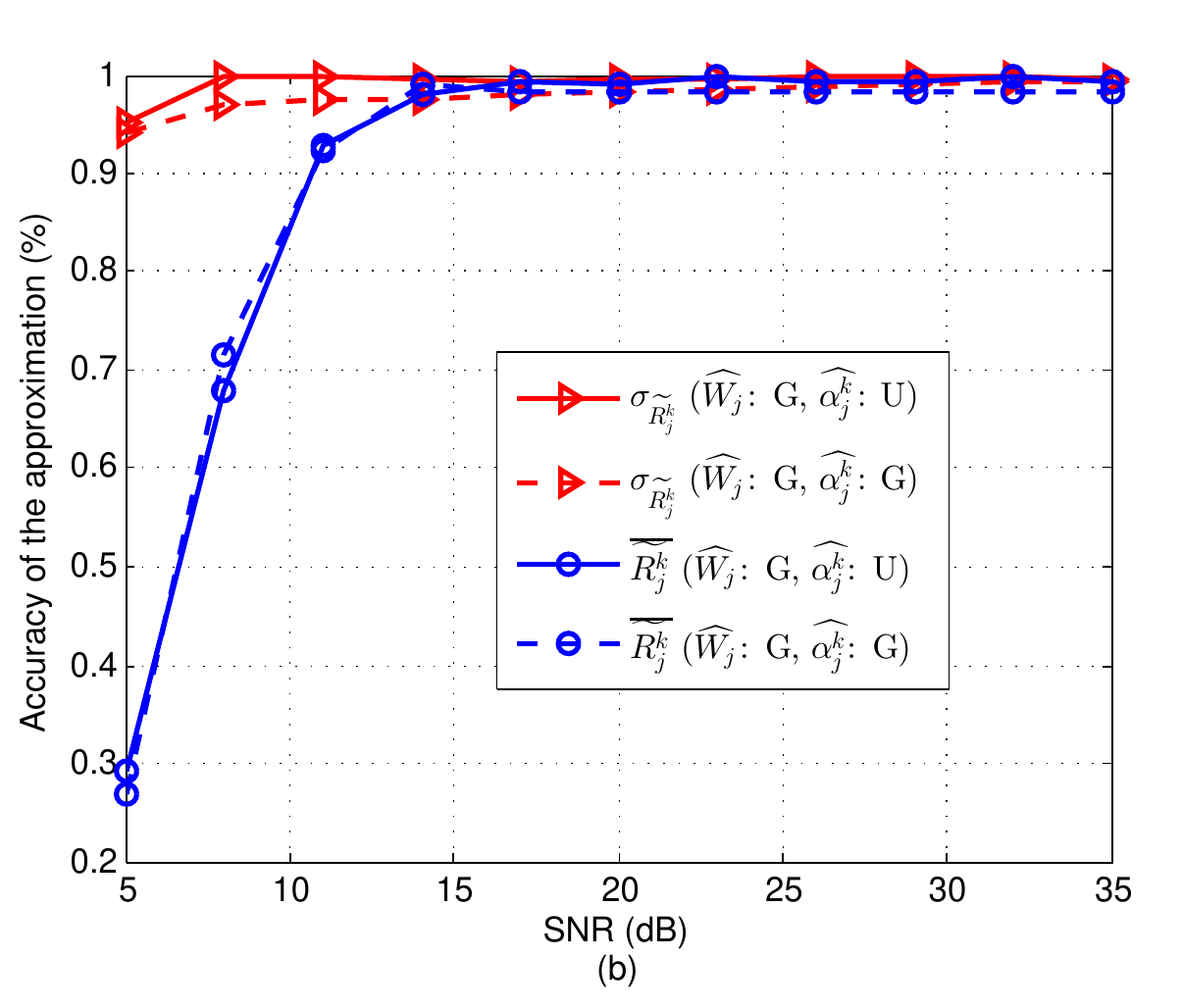}}
	\end{minipage}	\\

	\begin{minipage}[t]{0.5\linewidth}	
		\subfigure[Impact of the CV of $\widehat{W_j}$ and $\widehat{\alpha_j^k}$ on the CV of $\widehat{R_j^k}$.]{
			\includegraphics[width=\textwidth]{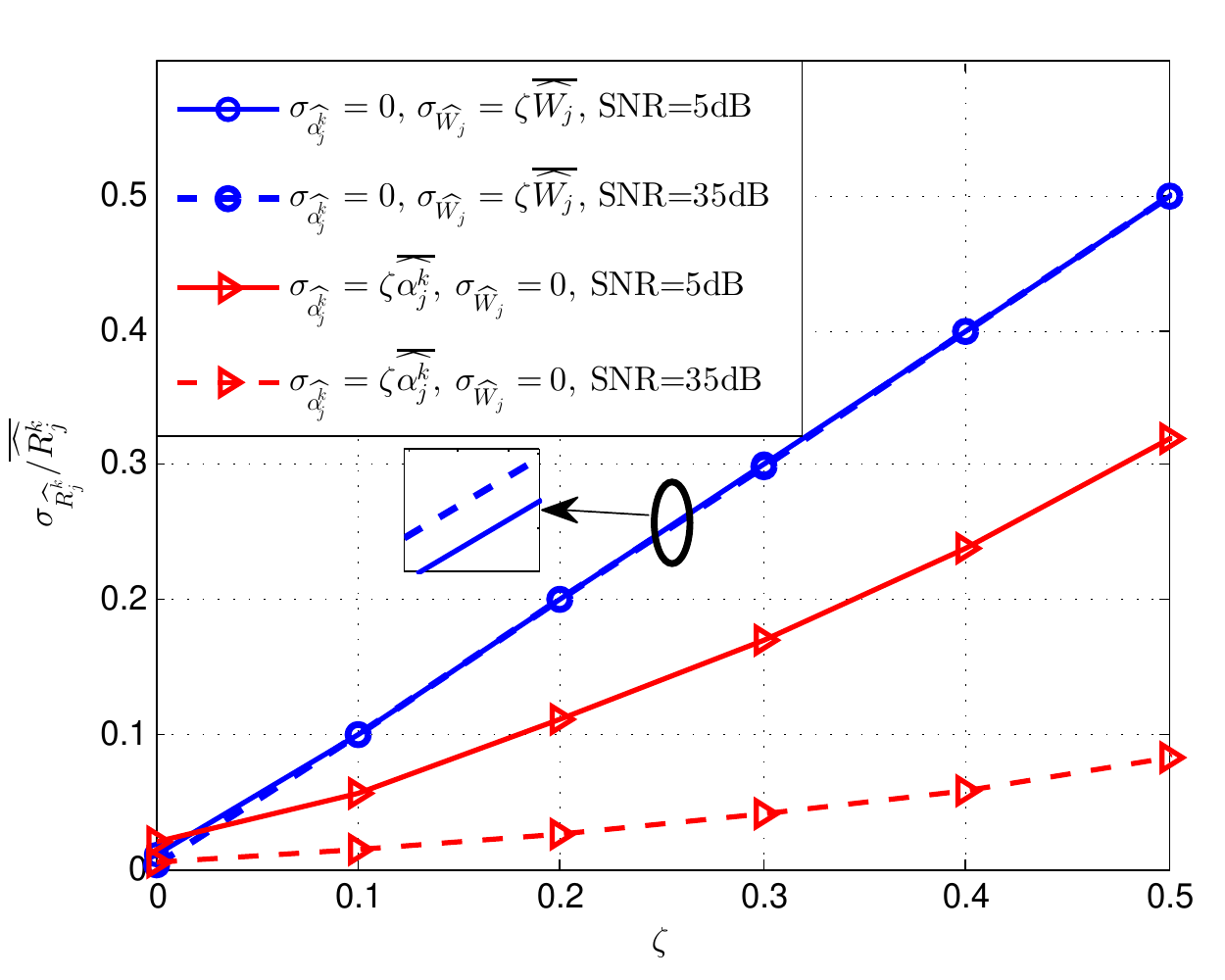}}
	\end{minipage}
	\begin{minipage}[t]{0.5\linewidth}	
		\subfigure[Impact of the bias of $\widehat{W_j}$ and $\widehat{\alpha_j^k}$ on the bias of $\widehat{R_j^k}$.]{
		\includegraphics[width=\textwidth]{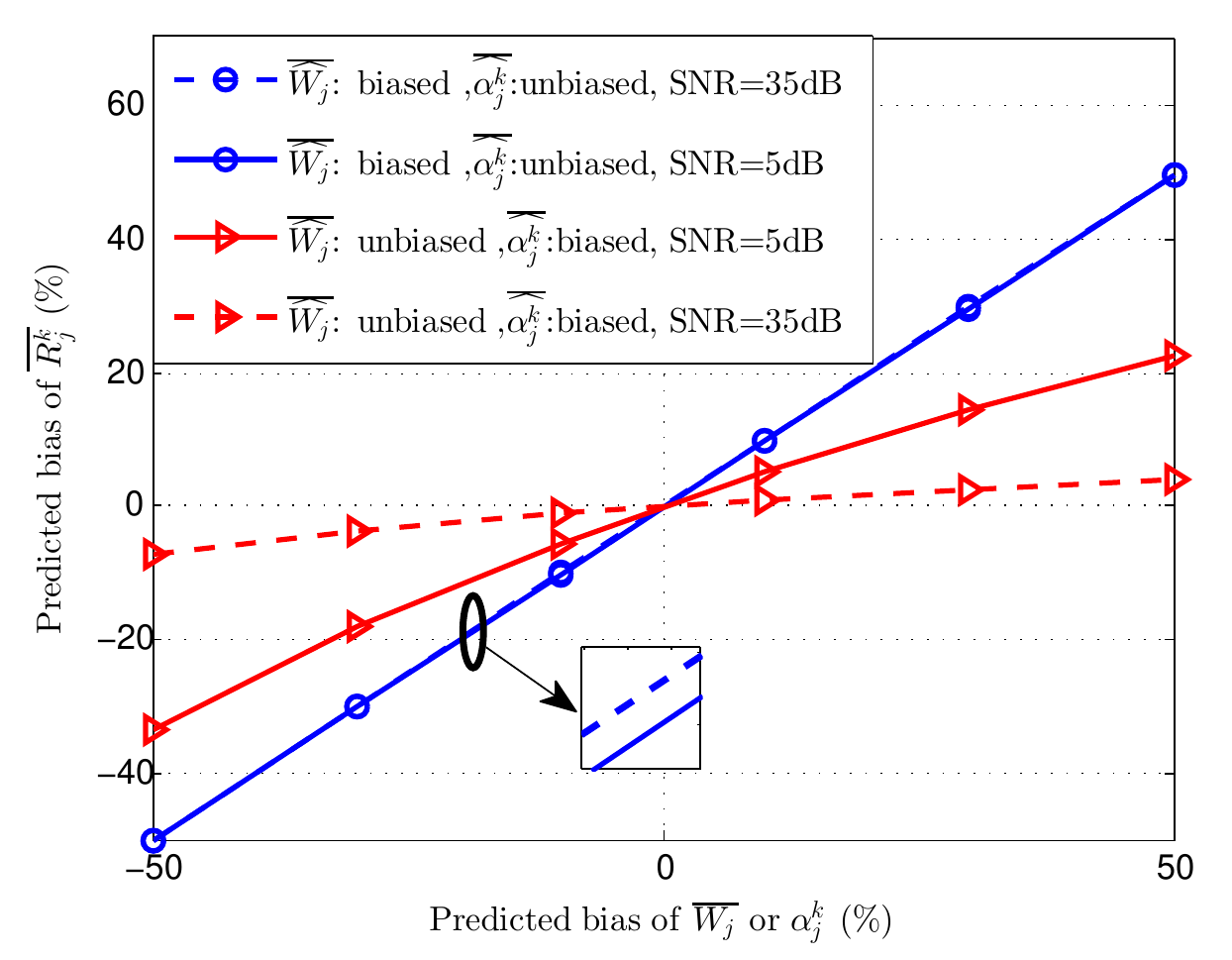}}
	\end{minipage}
	\caption{Validating the proposition. In the legends, ``G" and ``U" stand for Gaussian and uniform distributions, respectively.}\label{PPproof}
\end{figure}

 Fig. \ref{PPproof}(a) provides simulation and numerical results for the probability density function (PDF) of $\widetilde{R_j^k}$ when $\widehat{W_j}$ and $\widehat{\alpha_j^k}$ follow Gaussian and/or uniform distribution (the results have been normalized to have zero mean and unit variance for easy comparison). The average SNR is set as 5 dB or 35 dB, which represents the SNR when the MS is located at the cell edge or is closest to the BS when the MS moves along a straight line across the cell. Fig. \ref{PPproof}(b) shows the accuracy of the approximations used in \eqref{E:average_R} and \eqref{E:Var_R} when $\widehat{W_j}$ follows Gaussian distribution and $\widehat{\alpha_j^k}$ follows uniform or Gaussian distribution. Fig \ref{PPproof}(c) shows the impact of variance of
prediction errors of $\widehat{W_j}$ and $\widehat{\alpha_j^k}$ on the prediction error of $\widehat{R_j^k}$ when $\widehat{W_j}$ and $\widehat{\alpha_j^k}$ are unbiased predictions. To unity the units, the prediction error statistic is measured by coefficient of variation (CV, i.e., $\zeta=\sigma_{\widehat{W_j}}/\overline{\widehat{W_j}}$, taking residual bandwidth as an example). Fig. \ref{PPproof}(d) shows the impact of the prediction bias of $\overline{W_j}$ and $\alpha_j^k$ on the prediction bias of $\overline{R_j^k}$, where the prediction bias is normalized by true value, i.e., $(\overline{\widehat{W_j}}-\overline{W_j})/\overline{W_j}$, again using residual bandwidth as an example.
 	
	It is shown from Fig. \ref{PPproof}(a) that when $\widehat{W_j}$ follows Gaussian distribution, $\widetilde{R_j^k}$ is Gaussian as well, no matter what distribution $\widehat{\alpha_j^k}$ follows and under which SNR. However, when $\widehat{W_j}$ follows uniform distribution, $\widetilde{R_j^k}$ approximately follows uniform distribution. This suggests that the distribution of $\widetilde{R_j^k}$ mainly depends on that of $\widehat{W_j}$. It is shown from Fig. \ref{PPproof}(b)  that if $\widetilde{\alpha_j^k}$ follows Gaussian or uniform distribution, the approximations used in Proposition 1 are very accurate when the average SNR is larger than 15 dB. This implies that the relation between the prediction error statistics provided in the proposition are valid for predictive resource allocation, since its basic idea is to transmit at good channel condition \cite{Abou2013Predictive}. Fig. \ref{PPproof}(c) shows that the CV of $\widehat{W_j}$ has larger impact on the CV of $\widehat{R_j^k}$ compared to the CV of $\widehat{ \alpha_j^k}$. Fig. \ref{PPproof}(d) shows that when $\widehat{W_j}$ is with bias, the bias of $\widehat{R_j^k}$ grows linearly with the bias of $\widehat{W_j}$, while when $\widehat{\alpha_j^k}$ is with bias, the prediction bias of $\widehat{R_j^k}$ grows logarithmically with the prediction bias of $\widehat{ \alpha_j^k}$. This indicates that the variance and bias of $\widehat{W_j}$ have larger impact on those of $\widehat{R_j^k}$, which validates the proposition.

	\subsubsection{Performance gain brought by prediction}
	
\begin{figure}[!htb]
		\centering
		\vspace{-3mm}
		\includegraphics[width=.5\textwidth]{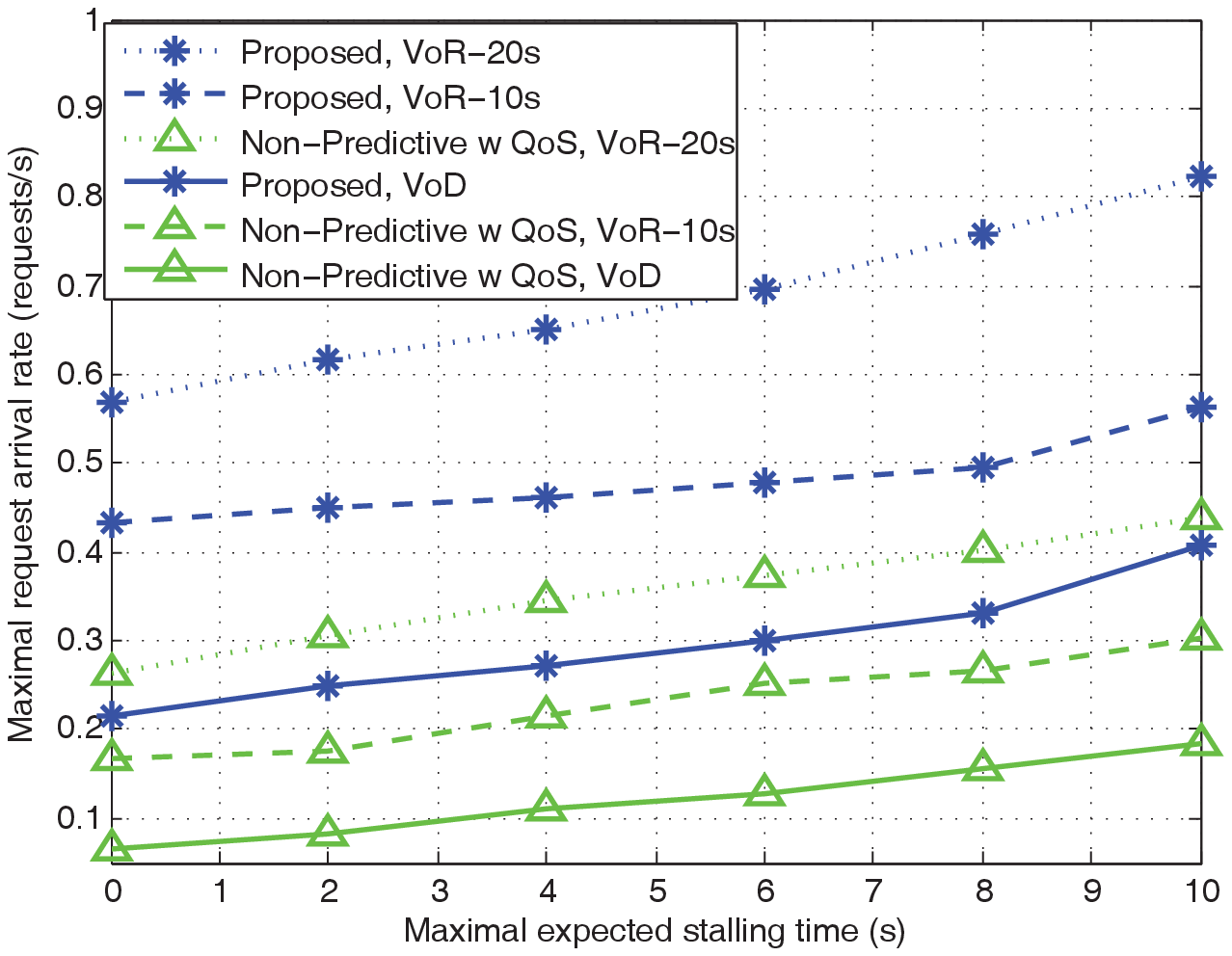}\\
		\caption{Gain from predicting average rate. ``VoR-20" or ``VoR-10'' in the legend means a VoR MS making reservation 20 s or 10 s in advance before the MS starts to play the video. }\label{PreVSnoPre}
	\end{figure}

	To demonstrate the gain from prediction, we compare ``Proposed" scheme with ``Non-predictive w QoS" scheme in Fig. \ref{PreVSnoPre}. Furthermore, by comparing the performance of serving VoD and VoR MSs with each scheme, we can observe the gain from knowing the contents to be transmitted  and the request arrival time in advance before the MSs initiate requests.
	
	
	By comparing ``Proposed" scheme with ``Non-predictive w QoS" scheme either when serving VoD or when serving VoR traffic, we can see remarkable gain from the prediction of future rate. By comparing the results obtained for VoR and VoD MSs with ``Proposed" scheme or  with ``Non-predictive w QoS" scheme, we can observe the additional gain of knowing the contents to be requested and the request arrival time, which is dramatic even with only 10 s reservation in advance. Moreover, the performance gap between these two schemes increases with the increase of reservation time. This indicates that the gain from predicting future rate will be even larger if the content to be requested and the request arrival time can be predicted.
	
	

	\subsubsection{Impact of using CSI} Most of existing works of predictive resource allocation do not consider CSI  both in optimization and in simulation, either by assuming that the small scale channel gain is static in each frame or by stating that its variation over time slots can be averaged out in a frame. However, the small scale channels of mobile users are impossible static, which in fact vary much faster than the large scale channels. On the other hand, despite that the variation of small scale channel gains among time slots in a frame can indeed  be averaged out when deriving the time-average rate of a frame if the gains are i.i.d., this does not mean that they can be ignored  during transmission. In practical cellular networks, CSI can be estimated at the BS by training at the start of each time slot. To help understand where the gain of our solution over existing works (as shown in the sequel) comes from, we compare the proposed scheme with ``Min-Time" scheme, both with or without using CSI, using the following way. When not using CSI during transmission in each time slot, both schemes schedule users sequentially. For example, if MS$_1$ and MS$_2$ need to download videos in the $j$th frame from a BS and the solution of problem ${\bf P1}$ is $s_j^1=0.4$ and $s_j^2=0.6$, then the BS will serve MS$_1$ in the first $40$ time slots in the $j$th frame and serve MS$_2$ in the remaining $60$ time slots. When using CSI, we use the transmission policy in section IV for both schemes after resource allocation plan are made by both schemes.

\begin{figure}[!htb]
	\centering
	\vspace{-3mm}
	\includegraphics[width=.5\textwidth]{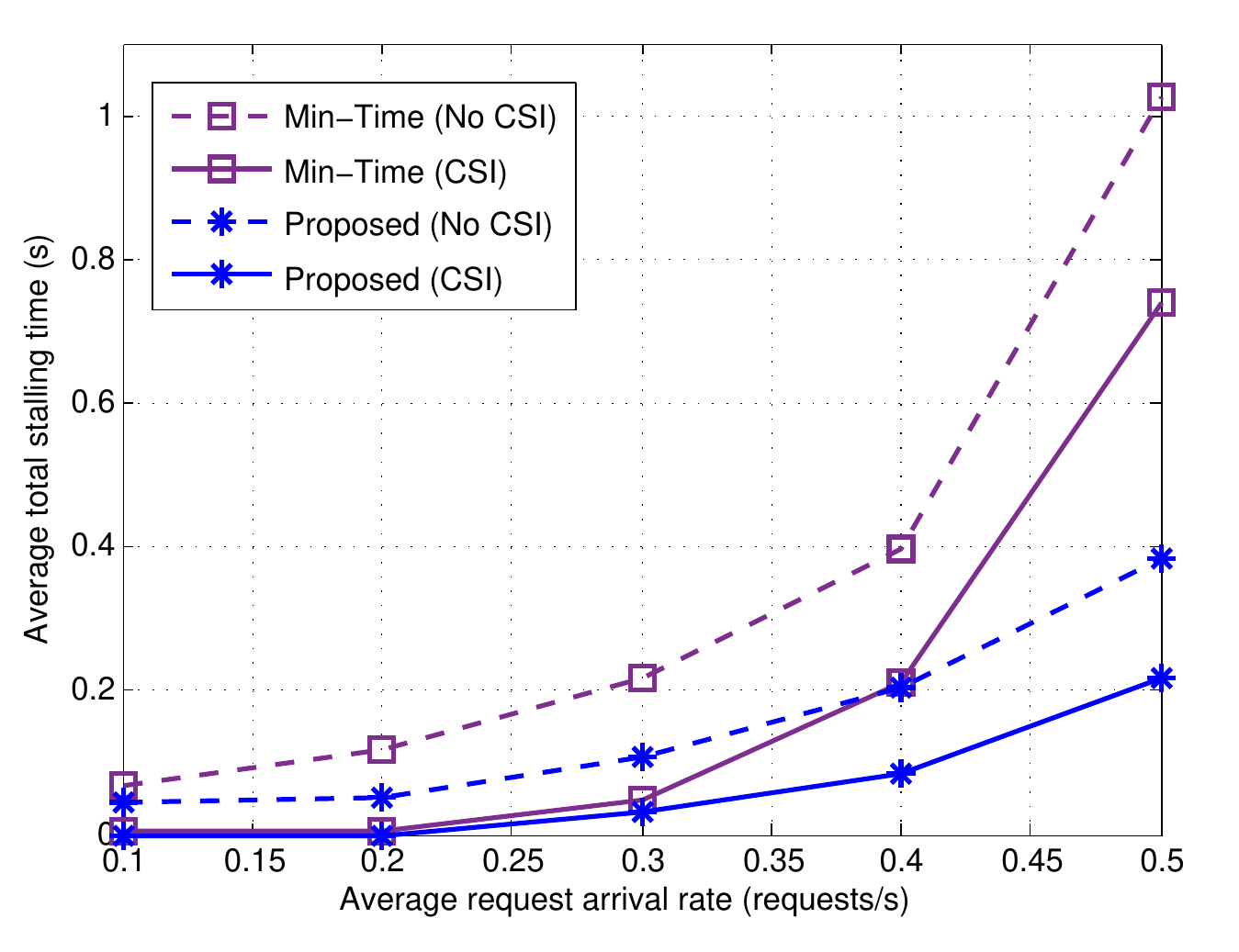}\\
	\caption{Impact of using CSI on the QoS of VoD MSs.}\label{CSI2}
	\end{figure}

	
	 Fig. \ref{CSI2} shows the average total stalling time versus average request arrival rate of the VoD MSs. We can observe the performance loss in QoS, especially when the average request arrival rate is high. Extensive simulations show that the schemes using CSI provide less stalling frequency than those without CSI, which are not shown for conciseness.

	
	\subsubsection{Comparison with other schemes} In what follows, we compare the performance of the proposed scheme with other schemes. In all the predictive schemes, $\alpha_j^k$ and $\overline{W_j}$ are predicted with errors modelled in subsection IV.A. For a fair comparison, the transmission policy in section IV is used for all predictive schemes to exploit the CSI available at each each time slot. To observe the impact of prediction errors, ``Proposed" scheme is also simulated when there are no prediction errors, i.e., $\sigma_{\widehat{W_j}}=\delta_j^k=0$.

\begin{figure}[!htb]
		\centering
		\vspace{-3mm}
		\includegraphics[width=.5\textwidth]{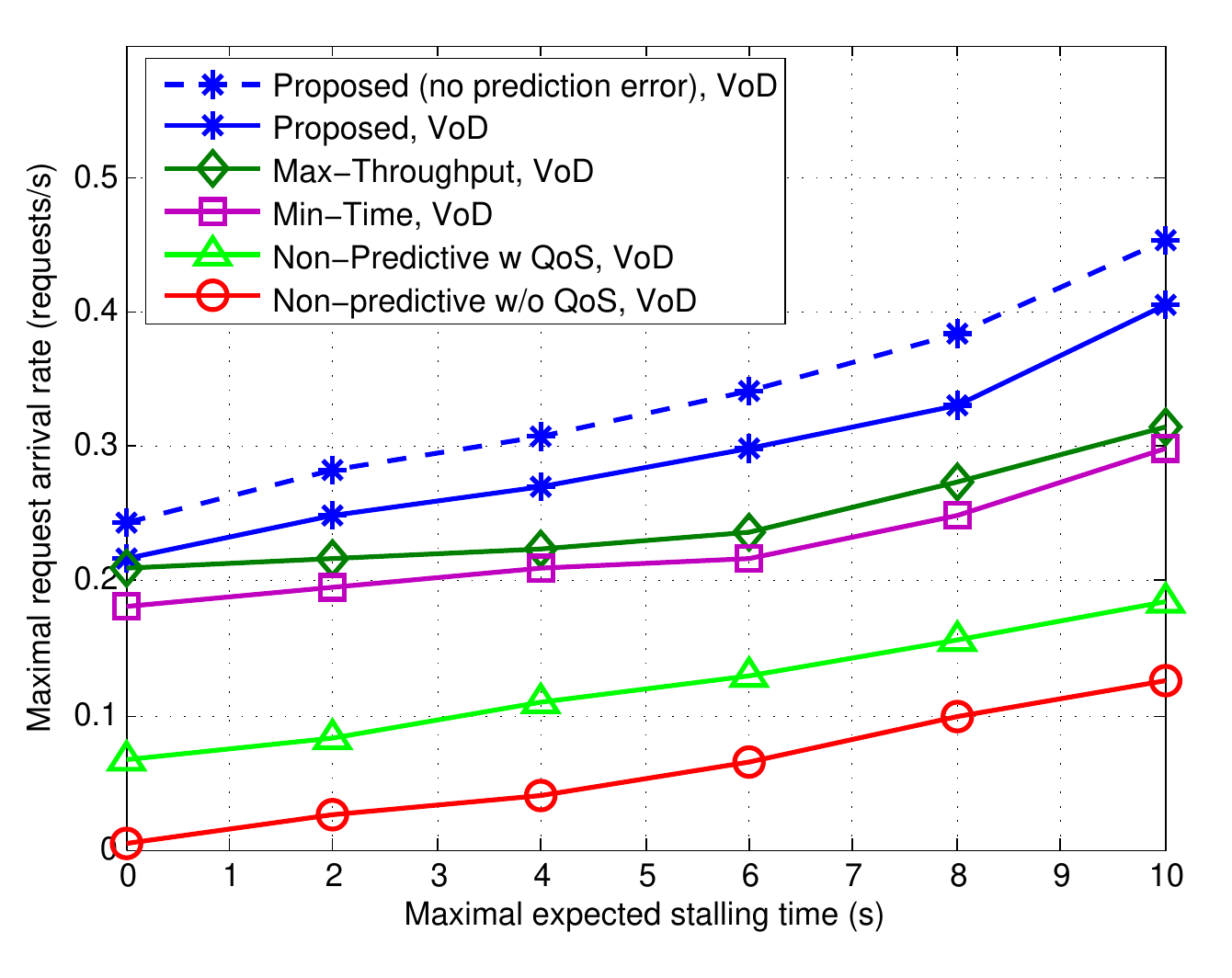}\\
		\caption{Performance comparison in terms of traffic carrying ability of the network with given tolerance of QoS of the MSs.}\label{aVSw}
	\end{figure}

In Fig. \ref{aVSw}, we show the maximal average request arrival rate of the VoD MSs versus the expected  maximal stalling time of each MS, which reflect the capability of supporting high throughput for VoD service by exploiting residual resources. It is shown that when the maximal stalling time is 10s, the gain of ``Proposed" over ``Non-predictive w/o QoS"  is 230\%, the gain over ``Non-predictive w QoS" is 110\%, the gain over ``Min-Time" is  33\%, and the gain over ``Max-Throughput" is 29\%. We can also see that the performance loss caused by prediction errors is 10\% when ``Proposed" scheme is adopted.

\begin{figure}[!htb]
		\centering
		\vspace{-3mm}
		\includegraphics[width=.5\textwidth]{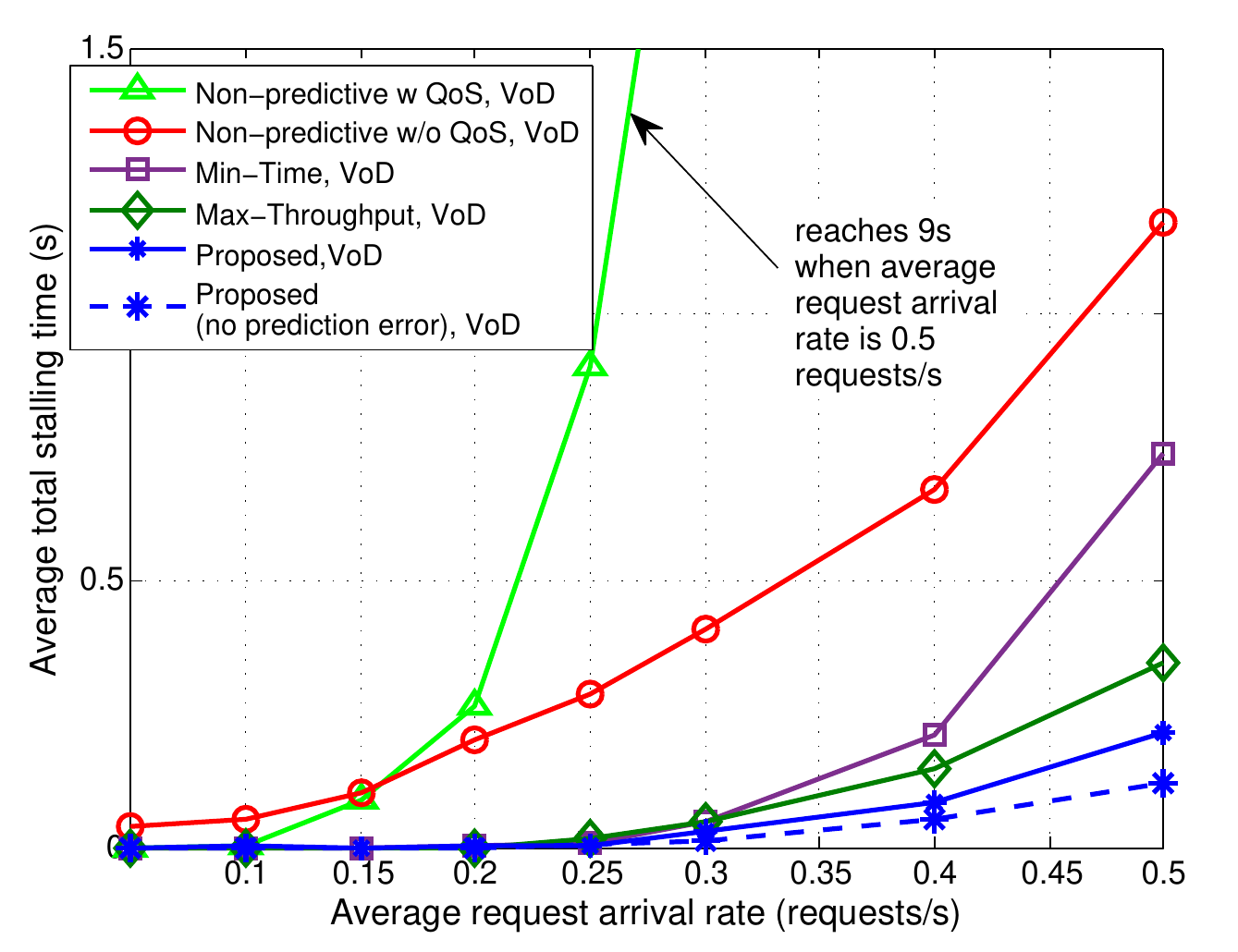}\\
		\caption{ Average QoS of the MSs with given traffic load of the MSs.}\label{wVSa}
	\end{figure}

In Fig. \ref{wVSa}, we show the average total stalling time of all the MSs versus the average request arrival rate of the MSs, which can reflect the average QoS of the MSs for a given traffic load. It is shown that when the average request arrival rate is 0.5 requests/s,  the gain of ``Proposed" over ``Non-predictive w QoS" in terms of reducing the average total stalling time is 98\%, the gain over ``Non-predictive w/o QoS" is 84\%, the gain over ``Min-Time" is 76\%, and the gain over ``Max-Throughput" is 43\%. We can also see  that the performance loss caused by prediction errors is 54\% when ``Proposed" scheme is used.

\begin{figure}[!htb]
		\centering
		\vspace{-3mm}
		\includegraphics[width=\textwidth]{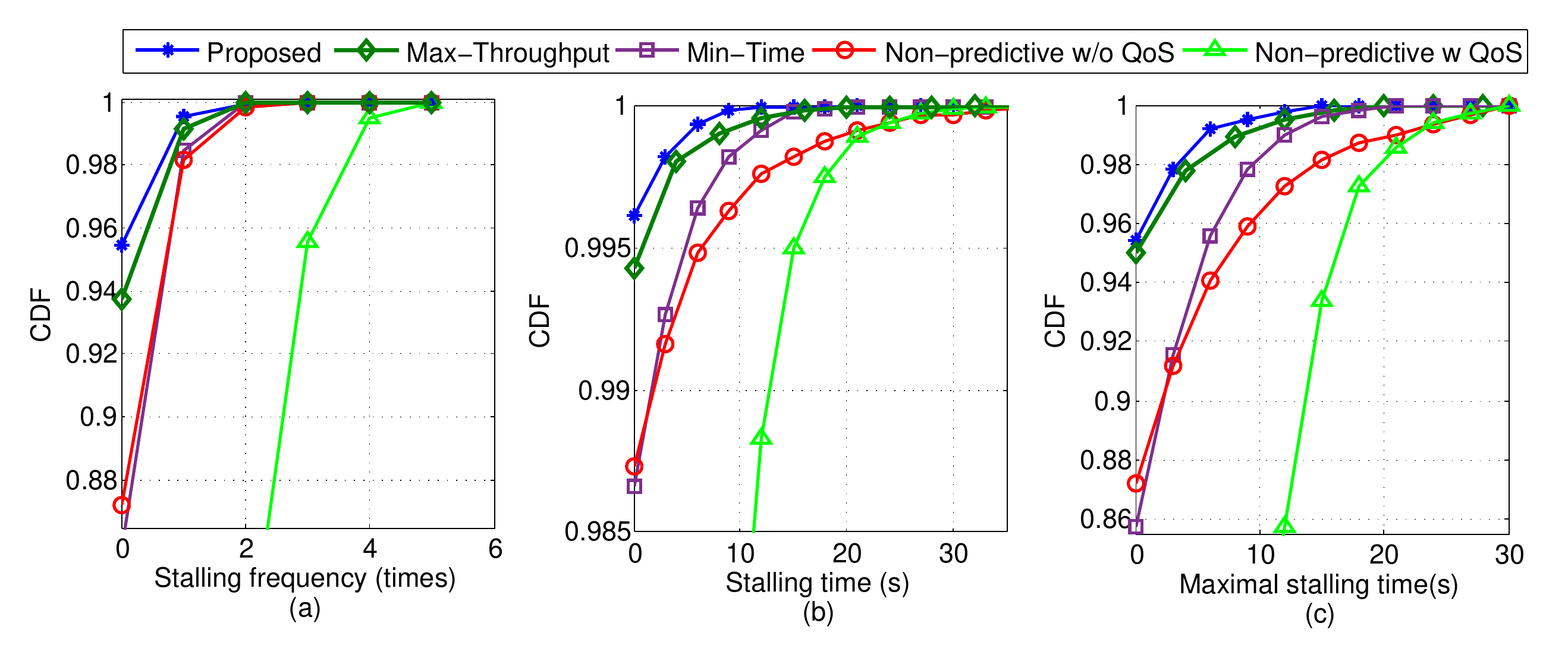}\\
		\caption{CDF of QoS-related key indicators: (a) stalling frequency; (b) stalling time; (c) maximal stalling time of each MS, where the average request arrival rate of VoD MSs is 0.5 requests/s.}\label{cdf}
	\end{figure}

In Fig. \ref{cdf}, we show the cumulative distribution function (CDF) of several key performance indicators to characterize the QoS of the VoD MSs when the average request arrival rate is 0.5 requests/s. As expected, ``Proposed" scheme can provide the lowest stalling frequency, stalling time, and maximal stalling time among all schemes.

	\section{Conclusions}
	In this paper, we investigated the potential of predictive resource allocation
in supporting high request arrival rate of VoD service by exploiting network residual resources. To this end, we formulated a problem to optimize resource allocation plan  with predicted time-average rate for  VoD
MSs with asynchronously arrived random requests, and found the optimal solution. In practice, the predicted time-average rate can be obtained from the predicted average residual bandwidth  at each BS and the predicted
average channel gain of each VoD MS. To gain useful insight for the accuracy of each type of prediction, we showed the relation of the mean values and variances between their prediction
errors. Analytical results showed that the average residual bandwidth should be predicted accurately in order to reduce the prediction error of average rate, while the average channel gain is unnecessary to predict with high accuracy. We developed a transmission policy according to the resource allocation plan where the instantaneous channel available at each time slot is used. Simulation and numerical results validated our analysis, and demonstrated that the proposed predictive resource allocation can support much higher traffic load than priori methods with given tolerance of QoS of the MSs. Besides,  the gain from prediction will be even more remarkable if the content to be requested and the request arrival time are able to be known only several seconds in advance.

\begin{appendices}
\numberwithin{equation}{section}

	\section{Proof of Proposition 1}
Since the proof is the same for all MSs, in the sequel we omit the superscript $k$ for notational simplicity.

i) We first show that $\widetilde{R_j}=\widehat{R_j}-R_j$ follows Gaussian distribution. Because $R_j = \frac{1}{T_s}\sum_{t=1}^{T_s}R_{j,t} =\frac{1}{T_s}\sum_{t=1}^{T_s} W_{j,t}\log_2(1+\frac{\alpha_j \|{\bf h}_{j,t}\|^2}{\sigma^2}P_{\max}) $, and $W_{j,t}$ and ${\bf h}_{j,t}$ are i.i.d. in all time slots within the $j$th frame, we have $\mathbb D\{R_j\} = \mathbb D\{R_{j,t}\}/T_s$. When $T_s\to\infty$, $\mathbb D\{R_j\}=0$, i.e., $R_j$ is deterministic. Then, the distribution of $\widetilde{R_j}$ depends on $\widehat{R_j}$. Hence, we only need to prove that $\widehat{R_j}\approx \widehat{W_j}\log_2\Big(\frac{\widehat{\alpha_j}N_t }{\sigma^2}P_{\max}\Big) \triangleq \widehat{W_j}\widehat{\gamma_j}$ follows Gaussian distribution.
	
	If $\overline{\widehat{\alpha_j}}-\delta_j/2 \leq \widehat {\alpha_j}\leq \overline{\widehat{\alpha_j}}+\delta_j/2$, the PDF of $\widehat{\gamma_j} \triangleq \log_2(\frac{\widehat{\alpha_j}N_t }{\sigma^2}P_{\max})$ can be expressed as \cite{zeidler2004oxford},
	\begin{eqnarray} \label{gamma}
	f_j^{(\gamma)}(\gamma) = \begin{cases}
	\displaystyle f_j^{(\alpha)}\Big(\frac{2^{\gamma}\sigma^2}{P_{\max}N_t}\Big)\frac{2^{\gamma}\sigma^2}{P_{\max}N_t}\ln 2,& \gm<\gamma<\gp, \\
	0, & \text{otherwise,}
	\end{cases}
	\end{eqnarray}
	where  $f_j^{(\alpha)}(\cdot)$ is the PDF of $\widehat{\alpha_j}$, $\gm=\log_2\Big(\frac{\overline{\widehat\alpha_j}-\frac{\delta_j}{2}}{\sigma^2}N_tP_{\max}\Big)$, $\gp = \log_2\Big(\frac{\overline{\widehat\alpha_j}+\frac{\delta_j}{2}}{\sigma^2}N_tP_{\max}\Big)$, and  $\overline{\widehat\alpha_j} = {\mathbb E}\{\widehat{\alpha_j}\}$.
	
	Then, the cumulative distribution function (CDF) of $\widehat{R_j}\approx \widehat{W_j}\widehat{\gamma_j}$ can be obtained as,
	\begin{eqnarray}\label{Rprob}
	F_j^{(R)}(r) &=& \Pr(R\leq r) \approx \int_0^{\infty}\Bigg(\int_{0}^{r/w}f_j^{(\gamma)}(\gamma){\rm d}\gamma\Bigg)f_j^{(W)}(w){\rm d}w \nonumber\\
	&=& \int_{0}^{r\big{/}\gp}\Bigg(\int_{\gm}^{r/w}f_j^{(\gamma)}(\gamma){\rm d}\gamma\Bigg)f_j^{(W)}(w){\rm d}w + \int_{r\big{/}\gp}^{r\big{/}\gm}\Bigg(\int_{\gm}^{r/w}f_j^{(\gamma)}(\gamma){\rm d}\gamma\Bigg)f_j^{(W)}(w){\rm d}w,\notag\\
	\end{eqnarray}
	where $f_j^{(W)}(\cdot)$ is the PDF of $\widehat{W_j}$.
	Since  $w<r\big{/}\gp$  in the first term, i.e., $r/w>\gp$, according to \eqref{gamma} the inner integral in the first term equals $1$. Similarly, since $w>r\big{/}\gp$ in the second term, i.e., $r/w<\gp$, the inner integral in the second term is less than $1$. Hence, $F_j^{(R)}(r)$ satisfies
	\begin{eqnarray}\label{Rconstraint}
	\int_{0}^{r\big{/}\gp}f_j^{(W)}(w){\rm d}w \leq F_j^{(R)}(r) \leq \int_{0}^{r\big{/}\gm}f_j^{(W)}(w){\rm d}w.
	\end{eqnarray}
	When $r\big{/}\gm-r\big{/}\gp\to 0$, the upper and lower bounds of $F_j^{(R)}(r)$ meet. This suggests that if $\widehat{W_j}$ follows Gaussian distribution, then $\widehat{R_j}$ and hence $\widetilde{R_j}$ also follow Gaussian distribution. From the definition of $\gm$ and $\gp$, the condition $r\big{/}\gm-r\big{/}\gp\to 0$ can be rewritten as
	\begin{eqnarray}\label{normalcondition}
	\frac{r\log_2\Big(\displaystyle\frac{\overline{\widehat\alpha_j}+\delta_j/2}{\overline{\widehat\alpha_j}-\delta_j/2}\Big)}{ \displaystyle\log_2\Big(\frac{\overline{\widehat\alpha_j}+{\delta_j}/2}{\sigma^2}N_tP_{\max}\Big)\displaystyle\log_2\Big(\frac{\overline{\widehat\alpha_j}-{\delta_j}/2}{\sigma^2}N_tP_{\max}\Big)}\to 0,
	\end{eqnarray}
	which holds when $\delta_j\ll\overline{\widehat\alpha_j}$ or $\frac{\overline{\widehat\alpha_j}}{\sigma^2}N_tP_{\max}$ is large.
	
ii) We then derive the mean value of the prediction error $\overline{\widetilde{R_j}}$. To this end, we derive the mean value of $\widehat{R_j}$ (denoted as $\overline{\widehat{R_j}}$) and the mean value of $R_j$ (denoted as $\overline{R_j}$). To derive $\overline{\widehat{R_j}}$, we first  derive the mean value of $\log_2\Big(1+\frac{\widehat{\alpha_j} \|{\bf{h}}_{j,t}\|^2}{\sigma^2}P_{\max}\Big)$, which is denoted as $\widehat{\mu_j}$.
Since $\widehat{\alpha_j}\thicksim\mathbb U(\mapha-\rou/2,\mapha+\rou/2)$ and the small scale channel is Rayleigh fading, it can be derived as,
	\begin{eqnarray}\label{R:1}
	\widehat{\mu_j}&=&\int_{-\infty}^{\infty}\int_{-\infty}^{\infty} \log_2\Big(1+\frac{\widehat{\alpha_j}\|{\bf{h}}_{j,t}\|^2}{\sigma^2}\Pw\Big)f_j^{(\alpha)}(\apha)f^{(H)}(\h)\dd\apha\dd\h \nonumber\\
&=& \int_{\mapha-\rou/2}^{\mapha+\rou/2}\left\{\int_{0}^{\infty} \log_2\Big(1+\frac{\widehat{\alpha_j}\|{\bf{h}}_{j,t}\|^2}{\sigma^2}\Pw\Big)f^{(H)}(\h)\dd\h\right\}\frac{1}{\delta_j}\dd\apha,
	\end{eqnarray}
where $f^{(H)}(\|{\bf{h}}\|^2)$ is the PDF of $\|{\bf{h}}_{j,t}\|^2$, which is
	\begin{eqnarray}\label{h2prob}
	f^{(H)}(\|{\bf{h}}\|^2)=\frac{(\|{\bf{h}}\|^2)^{N_t-1}e^{-\|{\bf{h}}\|^2}}{\Gamma(N_t)}.
	\end{eqnarray}

	When the predicted instantaneous SNR $\gg$ 1, $\log_2\Big(1+\frac{\widehat{\alpha_j} \|{\bf{h}}_{j,t}\|^2}{\sigma^2}\Pw\Big) \approx\log_2\Big(\frac{\widehat{\alpha_j} \|{\bf{h}}_{j,t}\|^2}{\sigma^2}\Pw\Big)$.
	After substituting \eqref{h2prob} and further considering the integral result,
	\begin{eqnarray}\label{result1}
	\int_0^{\infty} a\ln(bx)x^{N-1}e^{-cx} {\rm d}x \!\!\!&\overset{y=cx}{=}& \!\!\!ac^{-N}\ln\Big(\frac{b}{c}\Big)\int_0^{\infty}y^{N-1}e^{-y}\dd y + ac^{N}\int_0^{\infty} e^{-y} y^{N-1}\ln y\dd y \nonumber \\
	&\overset{(a)}{=}&\!\!\! ac^{-N}\Gamma(N)\left\{\ln\Big(\frac{b}{c}\Big)+\psi(N)\right\},\label{con2}
	\end{eqnarray}
	where $a>0$, $b>0$, $c>0$, $\Gamma(\cdot)$ is the Euler gamma function and $\psi(\cdot)$ is the digamma function, (a) comes from $\int_0^{\infty}y^{N-1}e^{-y}\dd y=\Gamma(N)$, and $\int_0^{\infty} e^{-y} y^{N-1}\ln y\dd y=\Gamma(N)\psi(N)$ \cite{zeidler2004oxford}, \eqref{R:1} can be derived as,
	\begin{eqnarray}\label{P:mean1}
	\widehat{\mu_j}&\approx&\frac{1}{\delta_j\ln 2 }\int_{\mapha-\rou/2}^{\mapha+\rou/2} \Bigg(\ln\Big(\frac{\widehat{\alpha_j}}{\sigma^2}P_{\max}\Big)+\psi(N_t)\Bigg)\dd\apha\notag\\
	&=& \frac{1}{\rou\ln2}\Bigg(\mapha\ln\Big(\frac{\mapha+\rou/2}{\mapha-\rou/2}\Big)+\frac{\rou}{2}\ln\Big(\frac{(\mapha+\rou/2)(\mapha-\rou/2)\Pw^2}{\sigma^4}\Big)
	+\rou\big(\psi(N_t)-1\big)\Bigg). \notag \nonumber
	\end{eqnarray}
	Since the residual bandwidth is independent from small scale channels of the NRT users, the mean value of the predicted time-average rate can be obtained as,
	\begin{eqnarray}
	\overline{\widehat{R_j}}={\mathbb E}\{\widehat{W_j}\}{\mathbb E}\Big\{\frac{1}{T_s}\sum_{t=1}^{T_s}\log_2(1+\frac{\widehat{\alpha_j} \|{\bf{h}}_{j,t}\|^2}{\sigma^2}P_{\max})\Big\}\approx\overline{\widehat{W_j}}\widehat{\mu_j}. \label{meanRjk}
	\end{eqnarray}
	
Similarly, the mean value of the time-average rate can be derived as
\begin{eqnarray}
\overline{R_j} \approx \overline{W_j}\Bigg(\log_2\Big(\frac{{\alpha_j}}{\sigma^2}P_{\max}\Big)+\frac{\psi(N_t)}{\ln 2}\Bigg),\label{meanRjktr}
\end{eqnarray}
where $\overline{W_j}={\mathbb E}\{W_{j,t}\}$, and the approximation is accurate when the instantaneous SNR is large.

Therefore, we obtain the mean value of the prediction error as in \eqref{E:average_R} with $\widehat{\mu_j}$ as in \eqref{xib}.

iii) Next, we derive the variance of the prediction error. Since $R_j$ is deterministic when $T_s \to \infty$, we only need to derive $\mathbb D\{\widehat{R_j}\}$. We first derive the the variance of $\frac{1}{T_s}\sum_{t=1}^{T_s}\log_2(1+\frac{\widehat{\alpha_j} \|{\bf{h}}_{j,t}\|^2}{\sigma^2}P_{\max})$, which is denoted as $ \widehat{ \sigma_j}^2$.
	
Since the small scale channel gains are i.i.d. among the time slots in each frame, we have $\widehat{ \sigma_j}^2=\frac{1}{T_s^2}\sum_{t_1=1}^{T_s}\sum_{t_2=1}^{T_s}\widehat \sigma_{j,t_1t_2}^2$, where $\widehat \sigma_{j,t_1t_2}^2={\rm cov}\Big(\log_2(1+\frac{\widehat{\alpha_j} \|{\bf{h}}_{j,t_1}\|^2}{\sigma^2}P_{\max}),\log_2(1+\frac{\widehat{\alpha_j} \|{\bf{h}}_{j,t_2}\|^2}{\sigma^2}P_{\max})\Big)$, and ${\rm cov}$ stands for covariance.
When the predicted instantaneous SNR $\gg$ 1 and $\widehat{\alpha_j}\thicksim{\mathbb U}(\overline{\widehat{\alpha_j}}-\delta_j/2,\overline{\widehat{\alpha_j}}+\delta_j/2)$, we have
	\begin{eqnarray}
	\widehat \sigma_{j,tt}^2\!\!\!&\approx&\!\!\!\int_{\mapha-\rou/2}^{\mapha+\rou/2}\Big\{ \int_{0}^{\infty} \log_2^2\Big(\frac{\widehat{\alpha_j} \|{\bf{h}}_{j,t}\|^2}{\sigma^2}\Pw\Big) f^{(H)}(\|{\bf{h}}\|^2)\dd \h \Big\} f^{(\alpha)}(\alpha)\dd\apha - \widehat{\mu_j}^2 \notag\\
	\!\!\!&=&\!\!\!\int_{\mapha-\rou/2}^{\mapha+\rou/2}\Big\{ \int_{0}^{\infty} \log_2^2\Big(\frac{\widehat{\alpha_j} \|{\bf{h}}_{j,t}\|^2}{\sigma^2}\Pw\Big) \frac{(\|{\bf{h}}_{j,t}\|^2)^{N_t-1}e^{-\|{\bf{h}}_{j,t}\|^2}}{\Gamma(N_t)} \dd \h \Big\} \frac{1}{\delta_j}\dd\apha - \widehat{\mu_j}^2. \nonumber
	\end{eqnarray}
By using the following integral result similarly derived as in obtaining \eqref{con2},
	\begin{eqnarray}
	\int_0^{\infty} a\ln^2(bx)x^{N-1}e^{-cx}{\rm d}x=ac^{-N}\Gamma(N)\Bigg(\Big(\ln\big(\frac{b}{c}\big)+\psi(N)\Big)^2+\psi'(N)\Bigg),\label{con4}
	\end{eqnarray}
where $a>0$, $b>0$, $c>0$, we have
	\begin{eqnarray}\label{R:5}
	\widehat \sigma_{j,tt}^2\approx\frac{1}{\delta_j \ln^2 2}\int_{\mapha-\rou/2}^{\mapha+\rou/2}\left\{ \Bigg(\ln\Big(\frac{\widehat{\alpha_j}}{\sigma^2}P_{\max}\Big)+\psi(N_t)\Bigg)^2+\psi'(N_t) \right\} \frac{1}{\delta_j}\dd\apha - \widehat{\mu_j}^2.
	\end{eqnarray}
Using the integral of $\ln^2(ax)$ and $\ln(ax)$ in \cite{zeidler2004oxford}, \eqref{R:5} can be further derived as,
	\begin{eqnarray}
	\widehat \sigma_{j,tt}^2&\approx &\frac{1}{\rou^2\ln^22}\left\{\mapha\rou\ln\Big(\frac{\mapha+\rou/2}{\mapha-\rou/2}\Big)\ln\Big(\frac{\Pw^2(\mapha-\rou/2)(\mapha+\rou/2)}{\sigma^4		 }\Big)\right.\notag\\
	&&~~~~~~~~~~+\big(\psi(N_t)-1\big)\Bigg(\rou^2\ln\Big(\frac{\Pw^2(\mapha-\rou/2)(\mapha+\rou/2)}{\sigma^4}\Big)+\ln\Big(\frac{\mapha+\rou/2}{\mapha-\rou/2}\Big)\Bigg)\notag\\
	&&~~~~~~~~~~+\frac{\rou^2}{2}\Bigg(\ln^2\Big(\frac{\Pw(\mapha+\rou/2)}{\sigma^2}\Big)+\ln^2\Big(\frac{\Pw(\mapha-\rou/2)}{\sigma^2}\Big)\Bigg)\notag\\
	&&~~~~~~~~~~\left.+\rou^2\big(\psi^2(N_t)-2\psi(N_t)+\psi'(N_t)+2\big)\right\}-\widehat{\mu_j}^2\notag\\
	&=&\frac{1}{\rou^2\ln^22}\Bigg(\Big(\frac{\rou^2}{4}-\mapha^2\Big)\ln^2\Big(\frac{\mapha+\rou/2}{\mapha-\rou/2}\Big)+\rou^2\big(1+\psi'(N_t)\big)\Bigg). \label{vari}
	\end{eqnarray}

When $t_1\neq t_2$, we have
	\begin{eqnarray}
	\widehat \sigma_{j,t_1t_2}^2=\int_{\overline{\widehat\alpha_j}-\delta_j/2}^{\overline{\widehat\alpha_j}+\delta_j/2}\Big\{\!\!\!\!\!\!\!\!&&\int_0^{\infty} \log_2\Big(1+\frac{\widehat{\alpha_j} \|{\bf{h}}_{j,t_1}\|^2}{\sigma^2}P_{\max}\Big)f^{(H)}(\|{\bf{h}}_{1}\|^2){\rm d}\|{\bf{h}}_{1}\|^2\notag\\
	&&\int_0^{\infty} \log_2\Big(1+\frac{\widehat{\alpha_j} \|{\bf{h}}_{j,t_2}\|^2}{\sigma^2}P_{\max}\Big)f^{(H)}(\|{\bf{h}}_{2}\|^2){\rm d}\|{\bf{h}}_{2}\|^2\Big\}\frac{1}{\delta_j}{\rm d}\alpha-\widehat{\mu_j}^2.\notag \label{lambda:2}
	\end{eqnarray}
	When the predicted instantaneous SNR $\gg 1$, upon substituting \eqref{h2prob} and by applying \eqref{con2}, we can obtain
	\begin{eqnarray}
	\widehat \sigma_{j,t_1t_2}^2&\approx&\frac{1}{\delta_j\ln^2 2}\int_{\overline{\widehat\alpha_j}-\delta_j/2}^{\overline{\widehat\alpha_j}+\delta_j/2}\Big(\ln\Big(\frac{\widehat{\alpha_j}}{\sigma^2}P_{\max}\Big)+\psi(N_t)\Big)^2{\rm d}\alpha-\widehat{\mu_j}^2\notag\\
	& = &\frac{1}{\rou^2\ln^22}\Bigg(\Big(\frac{\rou^2}{4}-\mapha^2\Big)\ln^2\Big(\frac{\mapha+\rou/2}{\mapha-\rou/2}\Big)+\rou^2\Bigg).
	\end{eqnarray}
	Since $\log_2(1+\frac{\widehat{\alpha_j} \|{\bf{h}}_{j,t}\|^2}{\sigma^2}P_{\max})$ and $\|{\bf{h}}_{j,t}\|^2$ are i.i.d. in all time slots in the $j$th frame, ${\widehat \sigma_{j,tt}^2}$ stays constant for any time slot $t$ and $\widehat \sigma_{j,t_1t_2}^2$ stays constant for any $t_1\neq t_2$ in the frame. Then, we have
	\begin{eqnarray}
	\widehat{ \sigma_j}^2&=&\frac{1}{T_s^2}\sum_{t_1=1}^{T_s}\sum_{t_2=1}^{T_s}\widehat \sigma_{j,t_1t_2}^2\approx\frac{1}{T_s^2}\Big(T_s\widehat \sigma_{j,tt}^2+(T_s^2-T_s)\widehat \sigma_{j,t_1t_2}^2\Big)\notag\\
	&=&\frac{1}{\rou^2\ln^22}\Bigg(\Big(\frac{\rou^2}{4}-\mapha^2\Big)\ln^2\Big(\frac{\mapha+\rou/2}{\mapha-\rou/2}\Big)+\rou^2\Bigg)+\frac{\psi'(N_t)}{T_s\ln^22}\notag \\
	& \overset{T_s\to\infty}{=} & \frac{1}{\rou^2\ln^22}\Bigg(\Big(\frac{\rou^2}{4}-\mapha^2\Big)\ln^2\Big(\frac{\mapha+\rou/2}{\mapha-\rou/2}\Big)+\rou^2\Bigg).
	\end{eqnarray}
Then, the variance of $\widetilde{R_{j}}$ can be obtained as:
\begin{eqnarray}
\sigma_{\widetilde{R_j}}^2 = \mathbb D\{\widetilde{R_j}\} &=& {\mathbb E}\{\widehat{R_{j}}^2\}-{\mathbb E}\{\widehat{R_{j}}\}^2 \approx (\sigma_{\widehat{W_j}}^2+\overline{\widehat{W_{j}}}^2)(\widehat{\sigma_{j}}^2+\widehat{\mu_{j}}^2) - \overline{\widehat{R_{j}}}^2.
\end{eqnarray}

iv) Finally, we analyze the impact of prediction biases of residual bandwidth and large scale channel gain. It is easy to see that if $\widehat{W_{j}}$ and $\widehat{\alpha_{j}}$  are unbiased, then $\widehat{R_{j}}$ will be unbiased. In what follows, we separately show the impact of the biases of $\widehat{W_{j}}$ and $\widehat{\alpha_{j}}$.

\begin{enumerate}
	\item $\widehat{W_j}$ is biased and $\widehat{\alpha_j}$ is unbiased: $\overline{\widehat{W_j}} = \eta\overline{W_j}$ and $\overline{\widehat\alpha_j} = \alpha_j$, where $\eta>0$ is a factor reflecting how large the bias $\overline{\widehat{W_j}} -\overline{W_j}$ is (when $\eta=1$, the prediction is unbiased). Then, when $N_t$ is large, the bias of the predicted time-average rate can be derived from \eqref{meanRjk} and \eqref{meanRjktr} as,
	\begin{eqnarray}
	\overline{\widetilde{R_j}}
	&\approx& \frac{\overline{W_j}}{\ln 2}\Bigg(\eta\Big(\frac{\overline{\widehat\alpha_j}}{\delta_j}\ln\Big(\frac{\mapha+\rou/2}{\mapha-\rou/2}\Big)+\frac{1}{2}\ln\Big(\frac{(\mapha+\rou/2)(\mapha-\rou/2)\Pw^2}{\sigma^4}\Big)-1\Big)\nonumber\\
&-&\ln\Big(\frac{\alpha_j}{\sigma^2}P_{\max}\Big)\Bigg). \notag
	\end{eqnarray}
	When $\overline{\widehat\alpha_j}\gg\delta_j$, $\frac{\overline{\widehat\alpha_j}}{\delta_j}\ln\Big(\frac{\mapha+\rou/2}{\mapha-\rou/2}\Big)\approx 1$\footnote{Since $\lim_{x\to\infty} x\ln(1+\frac{1}{x})=1$ and $\lim_{x\to\infty} \ln(1+\frac{1}{x})=0$, when $\overline{\widehat\alpha_j}\gg\delta_j$, $\frac{\overline{\widehat\alpha_j}}{\delta_j}\gg 1$, then $\frac{\overline{\widehat\alpha_j}}{\delta_j}\ln\Big(\frac{\mapha+\rou/2}{\mapha-\rou/2}\Big)=\Big(\frac{\overline{\widehat\alpha_j}}{\delta_j}-\frac{1}{2}\Big)\ln\Big(1+\frac{1}{\overline{\widehat\alpha_j}/\delta_j-1/2}\Big)+\frac{1}{2}\ln\Big(1+\frac{1}{\overline{\widehat\alpha_j}/\delta_j-1/2}\Big)\approx 1 + 0 =1$.}  and $\frac{1}{2}\ln\Big(\frac{(\mapha+\rou/2)(\mapha-\rou/2)\Pw^2}{\sigma^4}\Big)\approx \ln\Big(\frac{\overline{\widehat\alpha_j}}{\sigma^2}P_{\max}\Big)$. Then, the bias of the predicted time-average rate can be approximately
connected with the bias of the predicted residual bandwidth as,
	\begin{eqnarray}\label{E3}
	|\overline{\widetilde{R_j}}| \approx \left|\frac{\overline{W_j}}{\ln 2}\Bigg(\eta\ln\Big(\frac{\overline{\widehat\alpha_j}}{\sigma^2}P_{\max}\Big)-\ln\Big(\frac{\alpha_j}{\sigma^2}P_{\max}\Big)\Bigg)\right|=\overline{W_j} \log_2\Big(\frac{\alpha_j}{\sigma^2}P_{\max}\Big)|\eta-1|.
	\end{eqnarray}

	\item $\widehat{W_j}$ is unbiased  and $\widehat{\alpha_j}$ is biased: $\overline{\widehat{W_j}} = \overline{W_j}$ and $\overline{\widehat\alpha_j} = \eta\alpha_j$, where $\eta>0$ is a factor reflecting how large the bias $\overline{\widehat\alpha_j}-\alpha_j$ is. Again, when $N_t$ is large, the bias of the predicted time-average rate can be derived from \eqref{meanRjk} and \eqref{meanRjktr} as,
	\begin{eqnarray}
	\overline{\widetilde{R_j}} &=& \overline{\widehat{R_j}}-\overline{R_j} \notag\\
	&\approx& \frac{\overline{W_j}}{\ln 2}\Bigg(\frac{\overline{\widehat\alpha_j}}{\delta_j}\ln\Big(\frac{\mapha+\rou/2}{\mapha-\rou/2}\Big)+\frac{1}{2}\ln\Big(\frac{(\mapha+\rou/2)(\mapha-\rou/2)\Pw^2}{\sigma^4}\Big)-1\nonumber\\
&-&\ln\Big(\frac{\alpha_j}{\sigma^2}P_{\max}\Big)\Bigg). \notag
	\end{eqnarray}
	Again, using $\frac{\overline{\widehat\alpha_j}}{\delta_j}\ln\Big(\frac{\mapha+\rou/2}{\mapha-\rou/2}\Big)\approx 1$ and $\frac{1}{2}\ln\Big(\frac{(\mapha+\rou/2)(\mapha-\rou/2)\Pw^2}{\sigma^4}\Big)\approx \ln\Big(\frac{\overline{\widehat\alpha_j}}{\sigma^2}P_{\max}\Big)$ when $\overline{\widehat\alpha_j}\gg\delta_j$, the bias of the predicted time-average rate can be approximately connected with the
bias of the predicted large scale channel gain as
	\begin{eqnarray}\label{P:mean12}
	\overline{\widetilde{R_j}} \approx \frac{\overline{W_j}}{\ln 2}\Bigg(\ln\Big(\frac{\eta\alpha_j}{\sigma^2}P_{\max}\Big)-\ln\Big(\frac{\alpha_j}{\sigma^2}P_{\max}\Big)\Bigg) = \overline{W_j} \log_2(\eta).
	\end{eqnarray}
	Since the approximation in \eqref{P:mean12} is accurate when $\frac{\overline{\widehat\alpha_j}}{\sigma^2}P_{\max}\gg 1$, i.e., $\frac{\eta\alpha_j}{\sigma^2}P_{\max}\gg 1$, $\eta$ should satisfy $\eta\gg\frac{\sigma^2}{\alpha_jP_{\max}}$. It is not hard to see that $\left|\frac{\eta-1}{\log_2(\eta)}\right|$ is a monotonically increasing function of $\eta$, hence the following inequality holds,
	\begin{eqnarray}\label{E1}
	\left|\frac{\log_2\Big(\frac{\alpha_j}{\sigma^2}P_{\max}\Big)(\eta-1)}{\log_2(\eta)}\right|\gg\left|1-\frac{\sigma^2}{\alpha_jP_{\max}}\right|.
	\end{eqnarray}
\end{enumerate}

	When $\frac{\overline{\widehat\alpha_j}}{\sigma^2}P_{\max}\gg 1$, $\frac{\sigma^2}{\overline{\widehat\alpha_j}P_{\max}}\approx 0$. Then, we can show the relationship between \eqref{E3} and \eqref{P:mean12} as
	\begin{eqnarray}\label{E2}
	|\overline{\widetilde{R_j}}| \approx \overline{W_j} |\log_2(\eta)| \ll \overline{W_j} \log_2\Big(\frac{\alpha_j}{\sigma^2}P_{\max}\Big)|\eta-1|,
	\end{eqnarray}
which means that the impact of the prediction bias of large scale channel gain is much smaller than that of residual bandwidth on the prediction bias of time-average rate.

\end{appendices}
	\bibliography{IEEEabrv,GJ}
\end{document}